\documentclass[%
 reprint,
 amsmath,amssymb,
 aps,
,superscriptaddress]{revtex4-2}

\usepackage{graphicx}% Include figure files
\usepackage{dcolumn}% Align table columns on decimal point
\usepackage{bm}% bold math
\usepackage{autobreak}
\usepackage{textcomp}
\usepackage{ulem}
\usepackage{color}
\usepackage{colortbl}
\usepackage{upgreek}
\definecolor{lgreen}{rgb}{0.88,1,0.88}
\definecolor{lblue}{rgb}{0.88,1,1}

\usepackage[colorlinks,
            citecolor=blue,
            linkcolor=black,
            anchorcolor=black,
            urlcolor=black]{hyperref}
\makeatletter
\renewcommand*{\@fnsymbol}[1]{\ensuremath{\ifcase#1\or \dagger\or *\or \ddagger\or
    \mathsection\or \mathparagraph\or \|\or **\or \dagger\dagger
    \or \ddagger\ddagger \else\@ctrerr\fi}}
\makeatother

\begin{document}

\preprint{APS/123-QED}

\title{Fractal superconducting nanowires detect infrared single photons with 84\% system detection efficiency, 1.02 polarization sensitivity, and 20.8~ps timing resolution}% Force line breaks with \\

\author{Yun Meng}
 \thanks{These authors contributed equally to this work.}
\author{Kai Zou}%
 \thanks{These authors contributed equally to this work.}
\author{Nan Hu}%
 \thanks{These authors contributed equally to this work.}
\affiliation{%
 School of Precision Instrument and Optoelectronic Engineering, Tianjin University, Tianjin 300072, China\\
}%
\affiliation{%
 Key Laboratory of Optoelectronic Information Science and Technology, Ministry of Education, Tianjin 300072, China\\
}%
\author{Liang Xu}%
\author{Xiaojian Lan}%
\affiliation{%
 School of Precision Instrument and Optoelectronic Engineering, Tianjin University, Tianjin 300072, China\\
}%
\affiliation{%
 Key Laboratory of Optoelectronic Information Science and Technology, Ministry of Education, Tianjin 300072, China\\
}%
\author{Stephan Steinhauer}%
\author{Samuel Gyger}%
\author{Val Zwiller}%
\affiliation{
 Department of Applied Physics, Royal Institute of Technology (KTH), SE-106 91 Stockholm, Sweden\\
}
\author{Xiaolong Hu}%
 \altaffiliation{xiaolonghu@tju.edu.cn}
\affiliation{%
 School of Precision Instrument and Optoelectronic Engineering, Tianjin University, Tianjin 300072, China\\
}%
\affiliation{%
 Key Laboratory of Optoelectronic Information Science and Technology, Ministry of Education, Tianjin 300072, China\\
}%

%\date{\today}% It is always \today, today,
             %  but any date may be explicitly specified

\begin{abstract}
The near-unity system detection efficiency (SDE) and excellent timing resolution of superconducting nanowire single-photon detectors (SNSPDs), combined with their other merits, have enabled many classical and quantum photonic applications. However, the prevalent design based on meandering nanowires makes SDE dependent on the polarization states of the incident photons; for unpolarized light, the major merit of high SDE would get compromised, which could be detrimental for photon-starved applications. Here, we create SNSPDs with an arced fractal geometry that almost completely eliminates this polarization dependence of the SDE, and we experimentally demonstrate 84$\pm$3$\%$ SDE, 1.02$^{+0.06}_{-0.02}$ polarization sensitivity at the wavelength of 1575~nm, and 20.8~ps timing jitter in a 0.1-W closed-cycle Gifford-McMahon cryocooler, at the base temperature of 2.0~K. This demonstration provides a novel, practical device structure of SNSPDs, allowing for operation in the visible, near-, and mid-infrared spectral ranges, and paves the way for polarization-insensitive single-photon detection with high SDE and high timing resolution.

%\begin{description}
%\item[Usage]
%Secondary publications and information retrieval purposes.
%\item[Structure]
%You may use the \texttt{description} environment to structure your abstract;
%use the optional argument of the \verb+\item+ command to give the category of each item. 
%\end{description}
\end{abstract}

%\keywords{Suggested keywords}%Use showkeys class option if keyword
                              %display desired
\maketitle

%\tableofcontents

\section*{INTRODUCTION}

Because of their near-unity system detection efficiency (SDE)~\cite{marsili2013detecting,zadeh2017single, you93, reddy_exceeding_2019, reddy_superconducting_2020,chang_detecting_2020,hu_detecting_2020}, low dark-count rate (DCR)~\cite{hochberg_detecting_2019}, high count rate~\cite{zhao_counting_nodate,munzberg2018superconducting, tao_high_2019,zhang_16-pixel_2019}, excellent timing resolution~\cite{korzh_demonstration_2020,esmaeil_zadeh_efficient_2020}, and broad working spectral range~\cite{marsili_efficient_2012,verma_towards_2019,chen_mid-infrared_2020,zhang_superconducting_2016}, superconducting nanowire single-photon detectors (SNSPDs)~\cite{goltsman_picosecond_2001} have been widely used in classical and quantum photonic applications~\cite{hadfield_single-photon_2009}, ranging from LiDAR~\cite{taylor2019photon}, detection of luminescence from singlet oxygen~\cite{gemmell2013singlet}, quantum key distribution (QKD)~\cite{yin2016measurement}, to quantum computing~\cite{wang2019boson, quantum2020}. Indeed, these detectors have become indispensable tools and enabling components in the systems requiring faint-light detection. However, the prevalent design based on meandering nanowires yields polarization-dependent SDE, which could be problematic if information is encoded in polarization states. In particular, when the polarization states of the photons are unknown, time-varying, or random, it may not be possible to rotate the polarization states to maximize SDE of SNSPDs. Therefore, the major merit of these detectors would get severely compromised; and this compromise could be detrimental for many photon-starved applications that stringently require high SDE. For photon-number-resolving detection or coincidence photon counting, the fidelity to resolve $n$ photons or the $n$-fold coincidence count rate scales with SDE$^{n}$~\cite{zou2020superconductingPR,zou2020superconductingPRApplied}, which quickly drops if SDE decreases. In a QKD system using SNSPDs, the  polarization-dependent mismatch of SDE makes the system vulnerable for quantum hacking~\cite{wei2019implementation}. Furthermore, the subtle trade-offs between SDE and timing resolution~\cite{meng_fractal_2020} make their simultaneous optimization challenging. Recently, 85$\%$ SDE at the wavelength of 915~nm and 7.7~ps device timing jitter~\cite{esmaeil_zadeh_efficient_2020}, and 98\% SDE at the wavelength of 1425~nm and 26~ps system timing jitter~\cite{chang_detecting_2020}, were demonstrated on meandering SNSPDs, but the SDEs were still polarization-dependent.

To address the issue of polarization dependence of SDE, several approaches have been proposed and demonstrated, including spiral SNSPDs~\cite{dorenbos_superconducting_2008,huang_spiral_2017}, two orthogonal side-by-side meanders~\cite{dorenbos_superconducting_2008}, double-layer orthogonal meanders~\cite{verma_three-dimensional_2012}, SNSPDs involving compensating high-index materials~\cite{xu_demonstration_2017, mukhtarova_polarization-insensitive_2018}, specially designed SNSPDs with low polarization dependence at a certain wavelength~\cite{verma_high-efficiency_2015, reddy_exceeding_2019}, and fractal SNSPDs~\cite{meng_fractal_2020, Gu_fractal_2015, chi_fractal_2018}. These demonstrations all have successfully reduced the polarization sensitivity (PS, the ratio of the polarization-maximum SDE, SDE$_{\textrm{max}}$, over the polarization-minimum SDE, SDE$_{\textrm{min}}$~\cite{PSnote}) of SNSPDs; however, none of them could simultaneously preserve other major merits, in particular, high SDE and excellent timing resolution. Among these demonstrations, amorphous SNSPDs, made of WSi or MoSi, have exhibited over 80\%~\cite{verma_three-dimensional_2012, verma_high-efficiency_2015} and even over 90\% SDE~\cite{reddy_exceeding_2019}, however, their timing jitter ranges from 76~ps to 465~ps; on the other hand, polycrystalline SNSPDs, made of NbN or NbTiN, have shown better timing resolution, however, so far, the highest SDE demonstrated on polycrystalline SNSPDs with low-PS designs is 60\%~\cite{meng_fractal_2020}, still significantly lower than the state-of-the-art SDE$_{\textrm{max}}$ of meandering SNSPDs, which is over 90\% demonstrated by several research groups~\cite{marsili2013detecting,zadeh2017single, you93, reddy_exceeding_2019, reddy_superconducting_2020,chang_detecting_2020,hu_detecting_2020}. Therefore, it remains an outstanding challenge how to boost the SDE of SNSPDs with low PS to the level comparable to the SDE$_{\textrm{max}}$ of their meandering counterparts while simultaneously optimizing the timing resolution. 

Although the geometry of the fractal SNSPDs \cite{chi_fractal_2018,meng_fractal_2020} eliminated the global orientation of the nanowire, and therefore, significantly reduced PS, it was also this geometry that brought the major obstacle for further enhancing SDE and timing resolution. The fractal design contains a plethora of U-turns and L-turns that may limit the switching current, $I_{\rm sw}$, due to the current-crowding effect~\cite{clem_geometry-dependent_2011}, which may further affect SDE and timing resolution. In the past, we demonstrated fractal superconducting nanowire avalanche photodetectors (SNAPs)~\cite{meng_fractal_2020} with 60\% SDE, 1.05 PS, and 45~ps timing jitter. However, it is still elusive whether this route, using fractal geometry, is a practical one to combine high SDE, low PS, and low timing jitter.

In this paper, we report on our design and demonstration of a fiber-coupled fractal SNAP, fully packaged in a 0.1-W  closed-cycle  Gifford-McMahon (GM) cryocooler with the base temperature of 2.0~K,  achieving 84$\pm$3$\%$ SDE, 1.02$^{+0.06}_{-0.02}$ PS at the wavelength of 1575~nm, and 20.8~ps timing jitter. An enabling innovation is that we used an arced fractal geometry~\cite{su_superconducting_2008} for the nanowire to successfully reduce the current-crowding effect and therefore, increased $I_{\rm sw}$ to a level comparable to that in the meandering structure with the same nanowire width, thickness, and fill factor, achieving saturated, near-unity internal quantum efficiency, \textit{P}$_{\textrm{r}}$. We integrated the arced fractal nanowire with an optical micro-cavity, supported by dielectric distributed Bragg structures, for enhancing the optical absorptance, \textit{A}, of the nanowire~\cite{you93, zichi_optimizing_2019,hu_detecting_2020, reddy_superconducting_2020}.

\begin{figure*}[htbp]
\includegraphics[width=\textwidth]{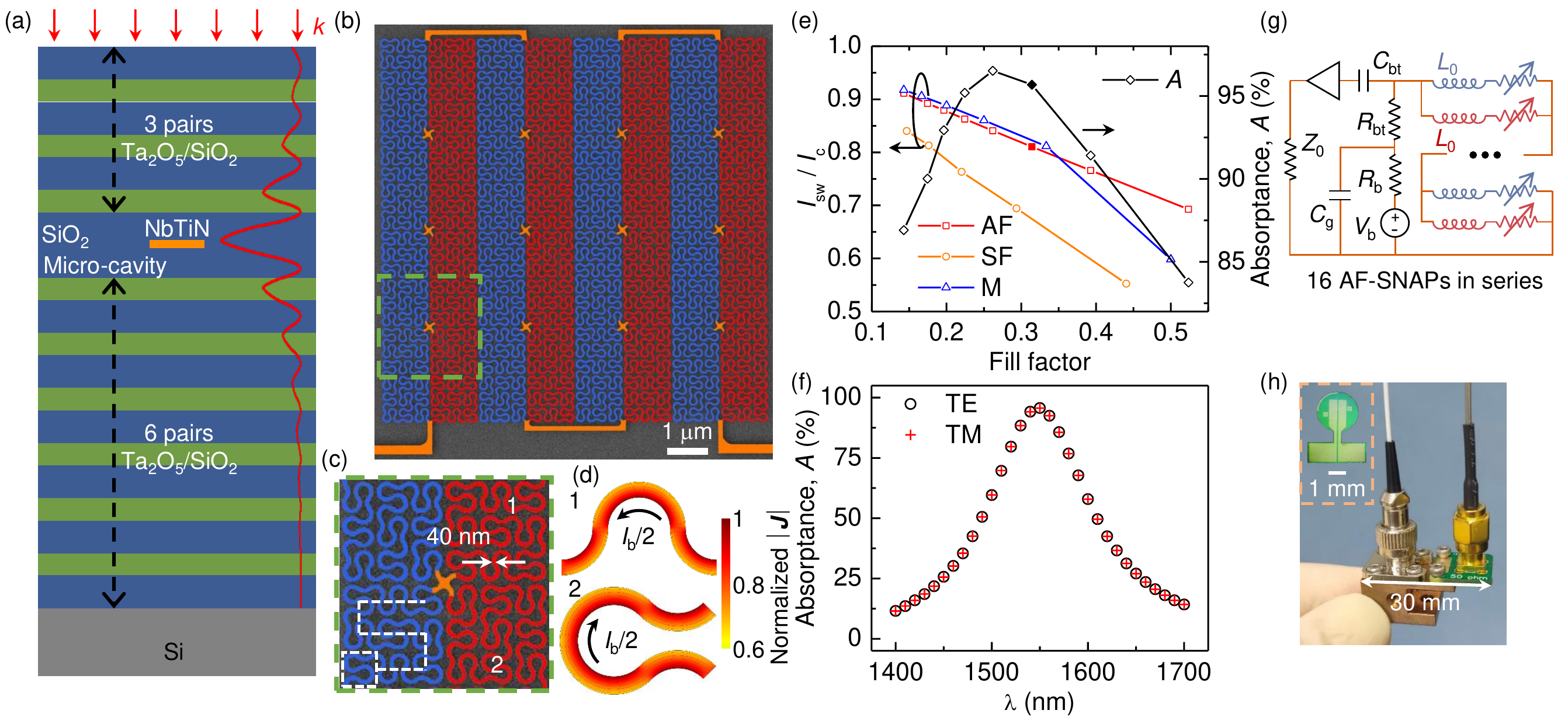}
\caption{\label{fig1}Design, fabrication, and packaging of arced fractal superconducting nanowire avalanche photodetectors (AF-SNAPs). (a) A schematic of the optical structure of an AF-SNAP. The nanowire is sandwiched in an optical micro-cavity supported by distributed Bragg structures, which are composed of dielectric alternating layers of silicon dioxide (SiO$_{2}$) and tantalum pentoxide (Ta$_2$O$_5$). The detector is illuminated from top and the red line presents the simulated distribution of the light intensity, assuming the absence of the nanowire. The nanowire is positioned in the micro-cavity where the light intensity is the strongest. (b) A false-colored scanning-electron micrograph of an AF-SNAP, in which the photosensitive nanowires are colored in red and blue, and the auxiliary structures are colored in orange. (c) A zoom-in micrograph of the region enclosed in the green-dashed box in (b). The width of the nanowire was measured to be 40~nm. (d) Simulated and normalized distribution of supercurrent density, $\lvert \textbf{\textit{J}} \lvert$, at the proximity of an L-turn and a U-turn, denoted in (c) as 1 and 2, respectively. (e) Simulated normalized switching currents, $I_{\rm sw}/I_{\rm c}$, of the arced fractal (red), standard fractal (orange), and meandering (blue) nanowires, as functions of the fill factor. The simulated optical absorptance (black) of the arced fractal nanowire is also presented as a function of the fill factor. The fill factor used in this work is 0.31. (f) Simulated optical absorptance of the AF-SNAP for two orthogonal linear polarization states, transverse-electric (TE) and transverse-magnetic (TM) states, as functions of the wavelength, $\lambda$. (g) Equivalent circuitry of the AF-SNAP, which is composed of 16 cascaded 2-SNAPs.  (h) A photograph of the chip package. Inset: a photograph of the keyhole-shaped chip.}
\end{figure*}

\section*{RESULTS AND DISCUSSION}

Figure 1 (a) presents a schematic of the optical structure of an arced fractal SNAP (AF-SNAP). Six pairs of alternating silicon dioxide (SiO$_{2}$) and tantalum pentoxide (Ta$_2$O$_5$) layers were deposited on a silicon substrate, functioning as the bottom Bragg reflector; three pairs formed the top reflector; in between a SiO$_2$ defect layer was sandwiched. The thicknesses of a SiO$_2$ layer and a Ta$_2$O$_5$ layer in the Bragg reflectors are 264~nm and 180~nm, respectively; and the thickness of the SiO$_2$ defect layer is 529~nm, targeting for the wavelength of 1550~nm with optimal optical absorptance. The red line in Fig.~\ref{fig1} (a) shows the simulated distribution of the light intensity in the dielectric stacks (without the nanowires) at the wavelength of 1550~nm for top illumination; and the NbTiN nanowires were designed to locate in the middle of the defect layer of the optical micro-cavity where the light intensity is the strongest. The thickness of the NbTiN film used in this work was 9~nm. Fig. 1 (b) presents a false-colored scanning-electron micrograph of a fabricated AF-SNAP, before the top Bragg layers were integrated. The photosensitive region of the detector was 10.2~µm by 10.2~µm, and the width of the nanowire was measured to be 40~nm [Fig. 1 (c)]. The photosensitive region of the detector is composed of 64 second-order arced fractal Peano curves~\cite{su_superconducting_2008} that are electrically connected according to the circuitry in Fig. 1 (g). In comparison to the the standard Peano fractal curve~\cite{chi_fractal_2018,meng_fractal_2020}, the arced Peano fractal curve~\cite{su_superconducting_2008} reduced the current-crowding effect at the turns. We simulated the normalized distribution of the supercurrent density, $\lvert \textbf{\textit{J}} \lvert$, at the proximity of an L-turn and a U-turn [Sec. S1 of Supporting Information (SI)], which are presented in Fig. 1 (d). Fig. 1 (e) further presents the simulated $I_{\textrm{sw}}$, normalized to the critical current of a straight nanowire with the same width and thickness, $I_{\textrm{c}}$, of the meandering, standard fractal, arced fractal nanowires, and the optical absorptance of the arced fractal nanowires, as functions of the fill factor. We used a commercial software COMSOL Multiphysics based on finite-element method for these simulations. At the fill factor of 0.31 used in this work, the simulated optical absorptance  for the plane wave at the wavelength of 1550~nm is 96$\%$, and the normalized switching current of the arced fractal nanowire is 0.81. As a comparison, with the same fill factor, 0.31, the normalized switching currents of the meandering nanowire and the standard fractal nanowire, are 0.82 and 0.67, respectively, further evidencing that the current-crowding effect in the arced fractal nanowire is significantly reduced, compared with that in the standard fractal one.  Detailed comparison of the distribution of the supercurrent density of these three types of geometry is presented in Sec. S1 of SI. Fig. 1 (f) presents the simulated optical absorptance, \textit{A}, of the AF-SNAP  for the plane wave as functions of wavelength for two orthogonal linear polarization states, denoted as transverse-electric (TE) and transverse-magnetic (TM) states. \textit{A} peaks at 1550~nm and remains above 50\% in the wavelength range from 1490~nm to 1610~nm. The simulation shows that \textit{A} is completely polarization-independent.  To investigate the coupling efficiency, $\upeta_{\rm c}$, we simulated the optical modes of the optical micro-cavity without the nanowires. The simulated mode-field diameter (MFD) at the plane of nanowires coupled with a Corning high-index optical fiber (HIF, HI 1060 FLEX) used in this paper was 6.8~\textmu m, which ensured the $\upeta_{\rm c}$ of 99\% assuming perfect alignment. In comparison, the MFD at the plane of nanowires coupled with a Corning SMF-28e+ optical fiber (SMF) was 10.7~\textmu m and the corresponding coupling efficiency is 89\%. The detector coupled with HIF is more tolerable to spatial misalignment than the detector coupled with SMF. Detailed simulation regarding optical modes of the cavity is presented in Sec. S2 of SI. Electrically, the detector was composed of sixteen cascaded 2-SNAPs, as we used previously~\cite{meng_fractal_2020}, and Fig. 1 (g) presents the equivalent circuit diagram. The chips were etched into the keyhole shape by Bosch process for self-aligned packaging~\cite{miller_compact_2011}. Detailed fabrication process is presented in Sec. S3 of SI. Fig. 1 (h) shows a photograph of the resulting chip package and the inset presents a photograph of a keyhole-shaped chip. In this package, the detector was self-aligned and directly coupled with a HIF, with a MFD of 6.3$\pm$0.3~\textmu m, which was connected to SMF, with a MFD of 10.4$\pm$0.5~\textmu m, through an in-line mode-field adapter (Sec. S4 of SI).

\begin{figure*}[htbp!]
\centering
\includegraphics[width=\linewidth]{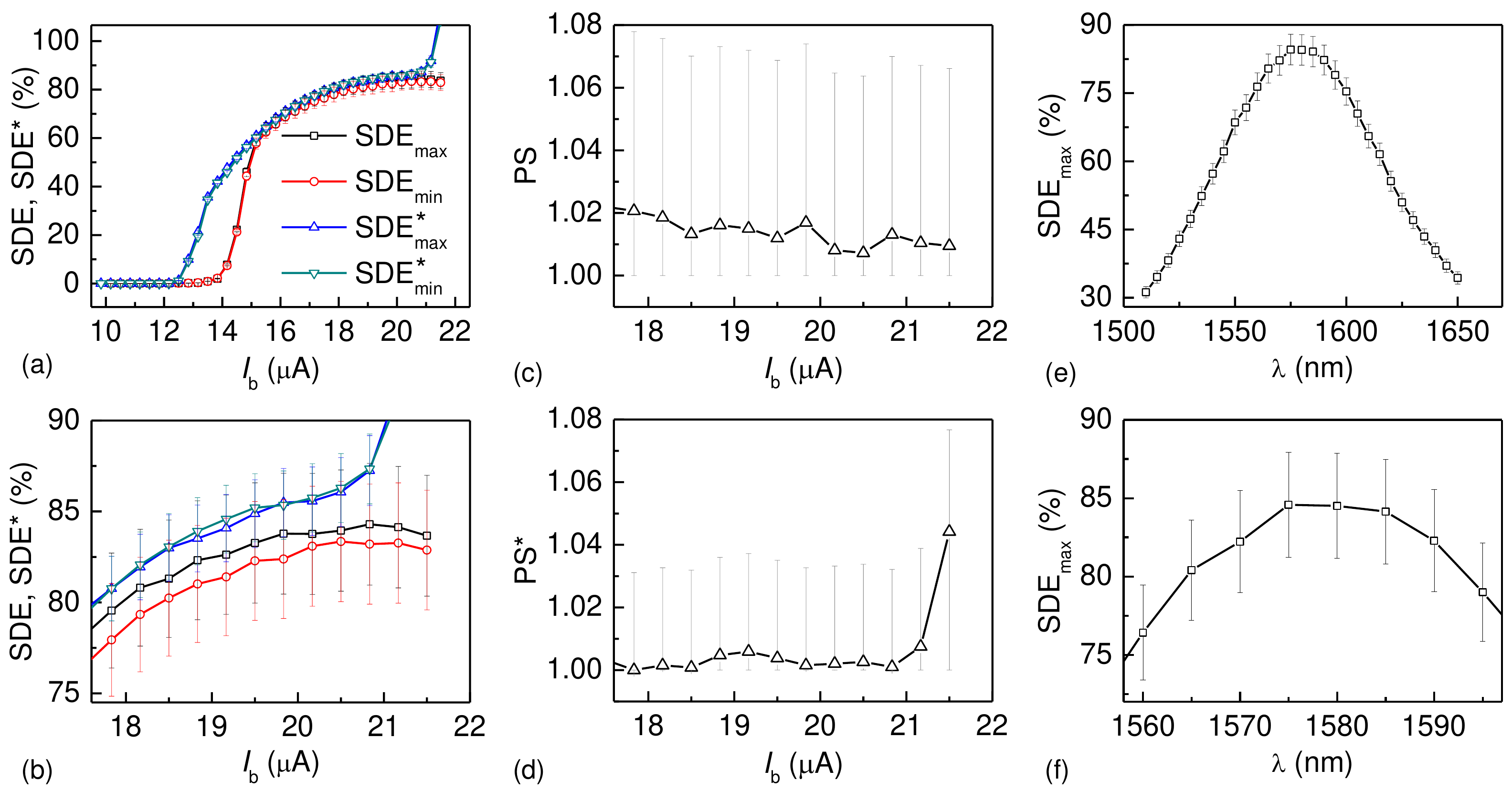}
\caption{\label{fig2}Measured system detection efficiency (SDE) and polarization dependence of an arced fractal superconducting nanowire avalanche photodetector (AF-SNAP). (a)~Measured $\rm SDE_{\rm max}$, $\rm SDE_{\rm min}$, $\rm SDE_{\rm max}^{*}$ and $\rm SDE_{\rm min}^{*}$ as functions of the bias current, $I_{\rm b}$. (b) Zoom-in view of (a) at the high-bias regime with $I_{\rm b}$ exceeding 17.5~\textmu A. (c)~Polarization sensitivity, PS, calculated from the measured $\rm SDE_{\rm max}$ and $\rm SDE_{\rm min}$. (d) ~Polarization sensitivity, PS$^{*}$, calculated from the measured $\rm SDE_{\rm max}^{*}$ and $\rm SDE_{\rm min}^{*}$.(e) Measured spectrum of $\rm SDE_{\rm max}$, at the bias current of 21.17~\textmu A. (f) Zoom-in view of the $\rm SDE_{\rm max}$ spectrum near the central wavelength of 1575~nm. }
\end{figure*}

We used the experimental setup, schematically presented in Sec. S4 of SI, to measure system detection efficiency and the polarization dependence. The base temperature for these measurements was 2.0~K. At this temperature, $I_{\rm sw}=21.67$~\textmu A. A cryogenic, low-noise microwave amplifier was mounted on the 40-K stage and used to amplify the output pulses. We first measured the DCR as a function of the bias current (Sec. S5 of SI). Then, we biased the detector at 21.17~\textmu A, tuned the laser wavelength and found that the system detection efficiency peaked at 1575~nm for this particular detector. The wavelength deviation from the designed wavelength with the maximum optical absorptance is presumably due to deviations of the thicknesses of the deposited dielectric layers and the refractive indices. We then fixed the wavelength at 1575~nm and scanned the polarization states of the input light over the Poincaré sphere, and found the polarization states corresponding to $\rm {SDE}_{max}^{*}$ and $\rm SDE_{min}^{*}$; at these two polarization states, we measured $\rm {SDE}_{max}^{*}$ and $\rm SDE_{min}^{*}$ as the functions of the bias current [Fig. 2 (a)]. To accurately measure the SDE*, we calibrated each optical attenuator at each polarization state and each wavelength for these measurements (Sec. S6 of SI). In the high-bias regime ($I_{\textrm{b}}>20.17\ $\textmu A), as shown in Fig. 2 (a) and (b), the $\rm {SDE}^{*}$-$I_{\rm b}$ curves go upward, showing additional false counts other than the dark counts and showing unrealistic $\rm {SDE}^{*}$. Similar observations have previously been reported on meandering SNAPs~\cite{miki2017stable} and also SNSPDs~\cite{chen2015dark}.  We re-measured the $\rm {SDE}_{max}$ and $\rm {SDE}_{min}$ using the method based on time-correlated photon counting \cite{chen2015dark} to exclude the false counts. Note that we use SDE* to refer to the system detection efficiency directly measured with the CW laser [Fig. S5 (a) for the experimental setup], excluding the dark counts; and we use SDE to refer to the system detection efficiency measured by time-correlated photon counting, excluding all false counts [Fig. S5 (b) for the experimental setup]. The values of SDE and SDE* in Fig. 2 (a), (b), (e), and (f) take into account the fiber end-facet reflection that occurred when we used the optical power meter to measure the optical power coming out from the fiber to avoid under-calculating the optical power delivered to the cryogenic AF-SNAP system and, therefore, to avoid over-calculating SDE and SDE*~\cite{chang_detecting_2020} (Sec. S7 of SI). In Fig. 2, the associated error bars present the uncertainties with 68\% confidence ($k=1$)~\cite{gerrits2019calibration} of the measurements (Sec. S8 in SI). At the bias current of 19.83~\textmu A, $\rm {SDE}_{max}$ was measured to be 84$\pm$3$\%$, $\rm {SDE}_{min}$ was measured to be 82$\pm$3$\%$, and the resulting PS ($\rm {SDE}_{max}$/$\rm {SDE}_{min}$) was 1.02$^{+0.06}_{-0.02}$.  $\rm {SDE}_{max}^{*}$ and $\rm {SDE}_{min}^{*}$ were measured to be both 85$\pm$2$\%$, and the resulting PS* was 1.00$^{+0.03}_{-0}$. At this bias current, DCR  and false-count rate (FCR) were measured to be $2.1\times 10^{3}$~cps and $2.2\times 10^{3}$~cps, respectively (Sec. S5 of SI). At the bias current of 17.83~\textmu A,
$\rm {SDE}_{max}$ and $\rm {SDE}_{min}$ decreased to 80$\pm$3$\%$ and 78$\pm$3$\%$, respectively; $\rm {SDE}_{max}^{*}$ and $\rm {SDE}_{min}^{*}$ decreased to both 81$\pm$2$\%$ [Fig. 2 (b)]; and the measured FCR and DCR at this bias current were 3.7$\times10^2$~cps and 3.0$\times10^2$~cps, respectively. In the low-bias regime, $I_{\textrm{b}}<$15.83~\textmu A, the detector was unstable~\cite{marsili2012afterpulsing}, generating multiple false pulses with low amplitudes after detecting one photon [Sec. S9 of SI], and resulting in the pronounced deviation of SDE* from SDE. PS and PS$^{*}$, as functions of the bias current, were calculated and presented in Fig. 2 (c) and (d). Fig. 2 (e) presents $\rm {SDE}_{max}$ as a function of the wavelength, $\uplambda$, at the bias current of 21.17~\textmu A. The full width at half maxima (FWHM) of the spectrum of $\rm SDE_{max}$ is 110~nm, which is slightly smaller than the FWHM, 120~nm, of the designed spectrum of the optical absorptance [Fig. 1 (f)]. Fig. 2 (f) presents a zoom-in view of the $\rm {SDE}_{max}$ for the wavelengths ranging from 1560~nm to 1595~nm, in which SDE$_{\textrm{max}}>$75\%. We note that here, the measurements of SDE and SDE* were performed at relatively low average input-photon rates going into the cryogenic AF-SNAP system, 1.79$\times10^5$~s$^{-1}$ and 1.82$\times10^5$~s$^{-1}$, respectively.

\begin{figure}[htbp]
\centering
\includegraphics[width=6.5cm]{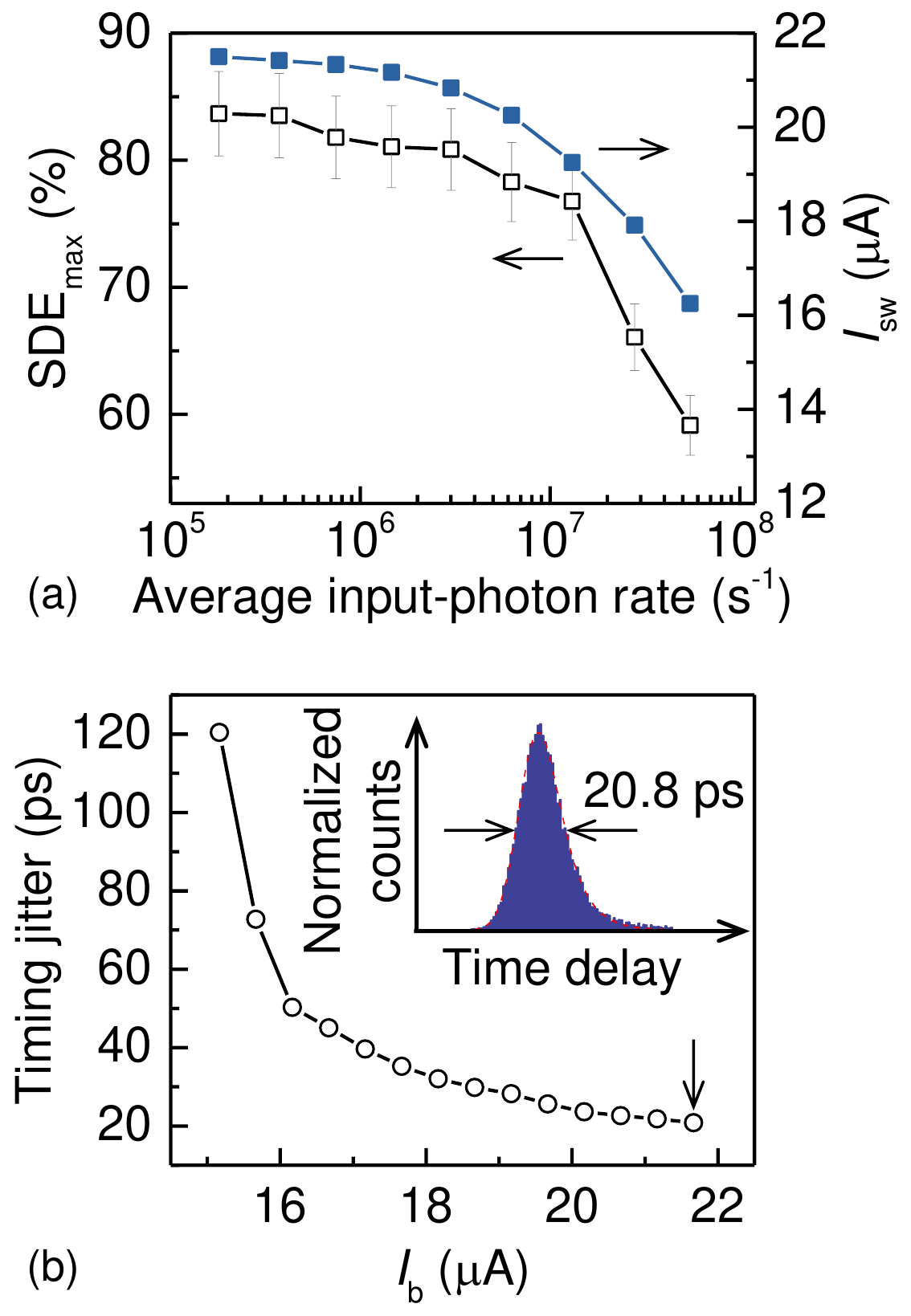}
\caption{\label{fig3}Timing properties of an arced fractal superconducting nanowire avalanche photodetector. (a) Measured $\rm SDE_{\rm max}$ and the switching current, $I_{\rm sw}$, as functions of the average input-photon rate. (b) Measured timing jitter as a function of the bias current, $I_{\rm b}$. Inset: measured time-delay histogram at the bias current of 21.67~\textmu A and the exponentially modified Gaussian (EMG) fitting, showing a full width at half maxima of the EMG fitting of 20.8~ps.}
\end{figure}

As the flux of the input photons increases, SDE would decrease. To characterize the SDE of the AF-SNAP at various input-photon rates, we measured the $\rm {SDE}_{max}$ as a function of the average input-photon rate (Sec. S10 of SI). Note that we used the oscilloscope  to measure the $\rm {SDE}_{max}$ in Fig.~3 (a) with the sampling rate of 2.5~Gsample/s and the time span per frame of 1~ms, because the maximum count rate of the input channel of the TAC is 12.5~Mcps, smaller than the photon-count rate we needed in this measurement. As the flux of the incident photon increases, the switching current decreases.  Each value of $\rm {SDE}_{max}$ in Fig.~3 (a) was measured at the bias current of 0.99$I_{\textrm{sw}}$.  The results show that when the average input-photon rate increases to 5.47$\times 10^7$~s$^{-1}$, $\rm {SDE}_{max}$ drops to 59\%, and the corresponding photon-count rate with FCR excluded was 26~Mcps.

Figure 3 (b) presents the measured timing jitter of the AF-SNAP by using a mode-locked fiber laser with the central wavelength of 1560~nm, a fast photodetector with 3-dB bandwidth of 40~GHz, and a real-time oscilloscope with bandwidth of 4~GHz. The experimental setup is schematically presented in Sec. S11 of SI. Each data point in Fig. 3 (b) is the FWHM of the exponentially modified Gaussian (EMG) fitting~\cite{sidorova2017physical} to the time-delay histograms [Fig. S12 (b)]. The lowest value of timing jitter was 20.8~ps at 21.67~\textmu A. The time-delay histogram and the EMG fitting is shown in the inset of Fig. 3 (b). Timing jitter monotonically increased with decreasing the bias current in the avalanche regime; for example, at $I_{\textrm{b}}=19.67$~\textmu A, timing jitter increased to 25.6~ps. As for other temporal properties, in the avalanche regime, the exponential fitting to the recovery edge of the output pulse shows a 1/$e$ time constant of 8.68~ns [Sec. S9 of SI].

We estimated the highest possible SDE for the current configuration of our system. The total transmittance of the two types of optical fibers connected through the mode-field adapter was measured to be 98\% at ambient temperature; the coupling efficiency, $\upeta_{\textrm{c}}$, between the high-index optical fiber and the photosensitive area was calculated to be 99\%, assuming perfect alignment; the optical absorptance, $A$, was simulated to be 96\% at the wavelength of 1550~nm; and the internal quantum efficiency, $P_{\rm{r}}$, was assumed to be 100\%. The product of these numbers gives an estimation of the highest possible SDE to be 93\%. The measured SDE, 84\%, is less than this estimation of the highest possible SDE presumably because (1) certain misalignment between the optical mode and the photosensitive area existed in the package and (2) $P_{\rm{r}}$ was less than 100\%.

\begin{figure*}[htbp]
\includegraphics[width=\linewidth]{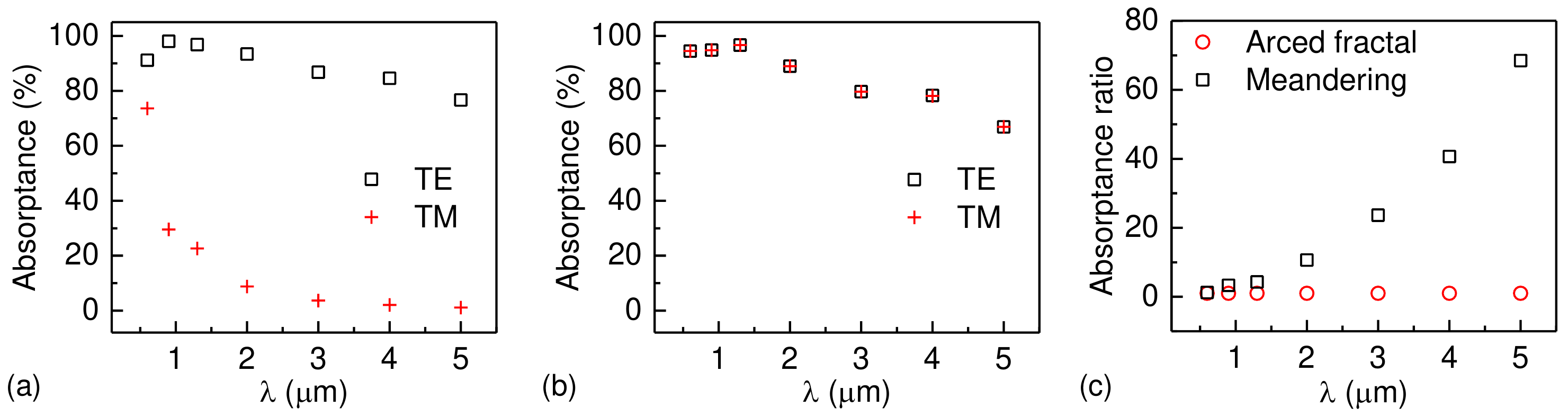}
\caption{\label{figS7} Simulated visible, near- and mid-infrared spectra of optical absorptance and the polarization dependence for the meandering and arced fractal SNSPDs. (a) Simulated optical absorptance of meandering SNSPDs for TE- and TM-polarization states. (b) Simulated optical absorptance of arced fractal SNSPDs for TE- and TM-polarization states. (c) Calculated absorptance ratios of TE- and TM-polarization states for meandering and arced fractal SNSPDs.}
\end{figure*}

The geometry of arced fractal nanowires can be applied to SNSPDs/SNAPs targeted for other interesting wavelengths by similarly re-designing the optical structures of the devices.  In particular, the polarization dependence of SDE becomes more severe at longer wavelengths for meandering SNSPDs, and we think that the geometry presented in this work would be useful for creating SNSPDs working in the mid-infrared with low PS. We simulated and optimized the optical absorptance of meandering [Fig. 4 (a)] and arced fractal SNSPDs [Fig. 4 (b)] at some additional wavelengths, 0.6~\textmu m, 0.9~\textmu m, 1.3~\textmu m, 2~\textmu m, 3~\textmu m, 4~\textmu m, and 5~\textmu m, for TE and TM polarization states. The optical structures are similar to that in Fig. 1 (a) except for that two, rather than three, pairs of alternating top layers maximize the optical absorptance of a meandering SNSPD for TE polarization and except for the modified thicknesses of the dielectric layers for different wavelengths. The simulation took into account the wavelength dependence of the refractive indices of the materials~\cite{palik1998handbook,hu2011efficient,bright2013infrared}, which are listed in Sec. S12 of SI. The calculated absorptance ratios of these two polarizations are presented in Fig. 4 (c). At the longer wavelengths, the absorptance ratio for the meandering SNSPD increases whereas the absorptance ratio for the arced fractal SNSPD remains constantly 1. As the polarization-dependent optical absorptance is the dominant contributor to the PS, and as this work demonstrates that the AF-SNAPs can reach high SDE and high timing resolution at the near infrared, we think that arced fractal SNSPDs/SNAPs should be good device structures for polarization-insensitive single-photon detection in the mid-infrared, as well as the visible, spectral ranges.

\section*{CONCLUSION}

In conclusion, we demonstrated a fiber-coupled AF-SNAP with 84$\pm$3$\%$ SDE, 1.02$^{+0.06}_{-0.02}$ residual PS at the wavelength of 1575~nm, and 20.8-ps timing jitter. The SDE was boosted to the level comparable to the amorphous SNSPDs/SNAPs with low PS~\cite{reddy_exceeding_2019, verma_three-dimensional_2012,verma_high-efficiency_2015}, but the timing resolution of the NbTiN AF-SNAP exceeded (See Sec. S13 of SI for the comparison; we also note that at their reported SDEs, amorphous SNSPDs/SNAPs listed in Sec. S13 of SI exhibited lower DCR than the FCR and DCR of the NbTiN AF-SNAP that we report in this paper, but amorphous SNSPDs/SNAPs require lower temperatures). These combined properties have not been achieved with any single-photon detectors reported previously and are enabled by our comprehensive device design. In particular, the arced fractal geometry of the nanowires is the key, enabling innovation that reduces the current-crowding effect and increases the switching current to the level comparable to that in the meandering SNSPDs and therefore, enhances both SDE and timing resolution. Since fractal SNSPDs were introduced in 2015~\cite{Gu_fractal_2015}, although we kept enhancing their performances~\cite{meng_fractal_2020, chi_fractal_2018}, it had been elusive whether the fractal designs of the nanowires could be practical device structures; it is this work that gives a positive and unambiguous answer by showing that fractal SNSPDs are practical devices with excellent comprehensive performances, comparable to meandering SNSPDs, on top of which low PS is added.  The arced fractal geometry is equally  applicable  to  designing  SNSPDs  working  in  other  spectral  ranges, in particular, mid infrared. This demonstration is a detector coupled with a single-mode optical fiber, but the same geometry can be used for detectors coupled with few- or multi-mode optical fibers and for detecting single photons coming from free space. Additionally, the negligibly small PS of the arced fractal SNSPDs/SNAPs would eliminate the security loophole, due to polarization-dependent mismatch of SDE, in the QKD systems~\cite{wei2019implementation}. We believe that this work paves the way for polarization-insensitive single-photon detection with high SDE and high timing resolution.

\begin{acknowledgments}
The authors would like to thank Prof. Qing-Yuan Zhao at Nanjing University and Dr. Jin Chang at TU Delft for helpful discussions. This work was supported by National Natural Science Foundation of China (NSFC) (62071322, 11527808, 61505141); National Key Research and Development Program of China (2019YFB2203600); Natural Science Foundation of Tianjin City (19JCYBJC16900).\\
\end{acknowledgments}

\section*{Author Contributions}
X. H., Y. M., and K. Z. conceived the project. Y. M., K. Z., N. H., and X. H. designed the devices and performed numerical simulation. S. S., S. G., and V. Z. sputtered NbTiN films. K. Z., N. H., and X. L. fabricated the devices. Y. M. and L. X. performed the measurements. X. H., Y. M., K. Z., and N. H. analyzed the data and wrote the paper. All authors commented and revised the paper. X. H. supervised the project.\\

\section*{COMPETING INTERESTS}
The authors declare no conflicts of interest.
%%%%%%%%%%%%%%%%%%%%%%%%%%%%%%%%   Yun Meng added  2020/05/08

\bibliography{sample}

%apsrev4-2.bst 2019-01-14 (MD) hand-edited version of apsrev4-1.bst
%Control: key (0)
%Control: author (8) initials jnrlst
%Control: editor formatted (1) identically to author
%Control: production of article title (0) allowed
%Control: page (0) single
%Control: year (1) truncated
%Control: production of eprint (0) enabled
\begin{thebibliography}{50}%
\makeatletter
\providecommand \@ifxundefined [1]{%
 \@ifx{#1\undefined}
}%
\providecommand \@ifnum [1]{%
 \ifnum #1\expandafter \@firstoftwo
 \else \expandafter \@secondoftwo
 \fi
}%
\providecommand \@ifx [1]{%
 \ifx #1\expandafter \@firstoftwo
 \else \expandafter \@secondoftwo
 \fi
}%
\providecommand \natexlab [1]{#1}%
\providecommand \enquote  [1]{``#1''}%
\providecommand \bibnamefont  [1]{#1}%
\providecommand \bibfnamefont [1]{#1}%
\providecommand \citenamefont [1]{#1}%
\providecommand \href@noop [0]{\@secondoftwo}%
\providecommand \href [0]{\begingroup \@sanitize@url \@href}%
\providecommand \@href[1]{\@@startlink{#1}\@@href}%
\providecommand \@@href[1]{\endgroup#1\@@endlink}%
\providecommand \@sanitize@url [0]{\catcode `\\12\catcode `\$12\catcode
  `\&12\catcode `\#12\catcode `\^12\catcode `\_12\catcode `\%12\relax}%
\providecommand \@@startlink[1]{}%
\providecommand \@@endlink[0]{}%
\providecommand \url  [0]{\begingroup\@sanitize@url \@url }%
\providecommand \@url [1]{\endgroup\@href {#1}{\urlprefix }}%
\providecommand \urlprefix  [0]{URL }%
\providecommand \Eprint [0]{\href }%
\providecommand \doibase [0]{https://doi.org/}%
\providecommand \selectlanguage [0]{\@gobble}%
\providecommand \bibinfo  [0]{\@secondoftwo}%
\providecommand \bibfield  [0]{\@secondoftwo}%
\providecommand \translation [1]{[#1]}%
\providecommand \BibitemOpen [0]{}%
\providecommand \bibitemStop [0]{}%
\providecommand \bibitemNoStop [0]{.\EOS\space}%
\providecommand \EOS [0]{\spacefactor3000\relax}%
\providecommand \BibitemShut  [1]{\csname bibitem#1\endcsname}%
\let\auto@bib@innerbib\@empty
%</preamble>
\bibitem [{\citenamefont {{Marsili}}\ \emph {et~al.}(2013)\citenamefont
  {{Marsili}}, \citenamefont {{Verma}}, \citenamefont {{Stern}}, \citenamefont
  {{Harrington}}, \citenamefont {{Lita}}, \citenamefont {{Gerrits}},
  \citenamefont {{Vayshenker}}, \citenamefont {{Baek}}, \citenamefont {{Shaw}},
  \citenamefont {{Mirin}},\ and\ \citenamefont {{Nam}}}]{marsili2013detecting}%
  \BibitemOpen
  \bibfield  {author} {\bibinfo {author} {\bibfnamefont {F.}~\bibnamefont
  {{Marsili}}}, \bibinfo {author} {\bibfnamefont {V.~B.}\ \bibnamefont
  {{Verma}}}, \bibinfo {author} {\bibfnamefont {J.~A.}\ \bibnamefont
  {{Stern}}}, \bibinfo {author} {\bibfnamefont {S.}~\bibnamefont
  {{Harrington}}}, \bibinfo {author} {\bibfnamefont {A.~E.}\ \bibnamefont
  {{Lita}}}, \bibinfo {author} {\bibfnamefont {T.}~\bibnamefont {{Gerrits}}},
  \bibinfo {author} {\bibfnamefont {I.}~\bibnamefont {{Vayshenker}}}, \bibinfo
  {author} {\bibfnamefont {B.}~\bibnamefont {{Baek}}}, \bibinfo {author}
  {\bibfnamefont {M.~D.}\ \bibnamefont {{Shaw}}}, \bibinfo {author}
  {\bibfnamefont {R.~P.}\ \bibnamefont {{Mirin}}},\ and\ \bibinfo {author}
  {\bibfnamefont {S.~W.}\ \bibnamefont {{Nam}}},\ }\bibfield  {title} {\bibinfo
  {title} {Detecting single infrared photons with 93\% system efficiency},\
  }\href@noop {} {\bibfield  {journal} {\bibinfo  {journal} {Nat. Photonics}\
  }\textbf {\bibinfo {volume} {7}},\ \bibinfo {pages} {210} (\bibinfo {year}
  {2013})}\BibitemShut {NoStop}%
\bibitem [{\citenamefont {{Esmaeil Zadeh}}\ \emph {et~al.}(2017)\citenamefont
  {{Esmaeil Zadeh}}, \citenamefont {{Los}}, \citenamefont {{Gourgues}},
  \citenamefont {{Steinmetz}}, \citenamefont {{Bulgarini}}, \citenamefont
  {{Dobrovolskiy}}, \citenamefont {{Zwiller}},\ and\ \citenamefont
  {{Dorenbos}}}]{zadeh2017single}%
  \BibitemOpen
  \bibfield  {author} {\bibinfo {author} {\bibfnamefont {I.}~\bibnamefont
  {{Esmaeil Zadeh}}}, \bibinfo {author} {\bibfnamefont {J.~W.~N.}\ \bibnamefont
  {{Los}}}, \bibinfo {author} {\bibfnamefont {R.~B.~M.}\ \bibnamefont
  {{Gourgues}}}, \bibinfo {author} {\bibfnamefont {V.}~\bibnamefont
  {{Steinmetz}}}, \bibinfo {author} {\bibfnamefont {G.}~\bibnamefont
  {{Bulgarini}}}, \bibinfo {author} {\bibfnamefont {S.~M.}\ \bibnamefont
  {{Dobrovolskiy}}}, \bibinfo {author} {\bibfnamefont {V.}~\bibnamefont
  {{Zwiller}}},\ and\ \bibinfo {author} {\bibfnamefont {S.~N.}\ \bibnamefont
  {{Dorenbos}}},\ }\bibfield  {title} {\bibinfo {title} {Single-photon
  detectors combining high efficiency, high detection rates, and ultra-high
  timing resolution},\ }\href@noop {} {\bibfield  {journal} {\bibinfo
  {journal} {APL Photonics}\ }\textbf {\bibinfo {volume} {2}},\ \bibinfo
  {pages} {111301} (\bibinfo {year} {2017})}\BibitemShut {NoStop}%
\bibitem [{\citenamefont {{Zhang}}\ \emph {et~al.}(2017)\citenamefont
  {{Zhang}}, \citenamefont {{You}}, \citenamefont {{Li}}, \citenamefont
  {{Huang}}, \citenamefont {{Lv}}, \citenamefont {{Zhang}}, \citenamefont
  {{Liu}}, \citenamefont {{Wu}}, \citenamefont {{Wang}},\ and\ \citenamefont
  {{Xie}}}]{you93}%
  \BibitemOpen
  \bibfield  {author} {\bibinfo {author} {\bibfnamefont {W.}~\bibnamefont
  {{Zhang}}}, \bibinfo {author} {\bibfnamefont {L.}~\bibnamefont {{You}}},
  \bibinfo {author} {\bibfnamefont {H.}~\bibnamefont {{Li}}}, \bibinfo {author}
  {\bibfnamefont {J.}~\bibnamefont {{Huang}}}, \bibinfo {author} {\bibfnamefont
  {C.}~\bibnamefont {{Lv}}}, \bibinfo {author} {\bibfnamefont {L.}~\bibnamefont
  {{Zhang}}}, \bibinfo {author} {\bibfnamefont {X.}~\bibnamefont {{Liu}}},
  \bibinfo {author} {\bibfnamefont {J.}~\bibnamefont {{Wu}}}, \bibinfo {author}
  {\bibfnamefont {Z.}~\bibnamefont {{Wang}}},\ and\ \bibinfo {author}
  {\bibfnamefont {X.}~\bibnamefont {{Xie}}},\ }\bibfield  {title} {\bibinfo
  {title} {Nbn superconducting nanowire single photon detector with efficiency
  over 90\% at 1550 nm wavelength operational at compact cryocooler
  temperature},\ }\href@noop {} {\bibfield  {journal} {\bibinfo  {journal}
  {Sci. China Phys. Mech. Astron.}\ }\textbf {\bibinfo {volume} {60}},\
  \bibinfo {pages} {120314} (\bibinfo {year} {2017})}\BibitemShut {NoStop}%
\bibitem [{\citenamefont {Reddy}\ \emph {et~al.}()\citenamefont {Reddy},
  \citenamefont {Nerem}, \citenamefont {Lita}, \citenamefont {Nam},
  \citenamefont {Mirin},\ and\ \citenamefont {Verma}}]{reddy_exceeding_2019}%
  \BibitemOpen
  \bibfield  {author} {\bibinfo {author} {\bibfnamefont {D.~V.}\ \bibnamefont
  {Reddy}}, \bibinfo {author} {\bibfnamefont {R.~R.}\ \bibnamefont {Nerem}},
  \bibinfo {author} {\bibfnamefont {A.~E.}\ \bibnamefont {Lita}}, \bibinfo
  {author} {\bibfnamefont {S.~W.}\ \bibnamefont {Nam}}, \bibinfo {author}
  {\bibfnamefont {R.~P.}\ \bibnamefont {Mirin}},\ and\ \bibinfo {author}
  {\bibfnamefont {V.~B.}\ \bibnamefont {Verma}},\ }\bibfield  {title} {\bibinfo
  {title} {Exceeding 95\% system efficiency within the telecom {C}-band in
  superconducting nanowire single photon detectors},\ }\href@noop {} {\bibfield
   {journal} {\bibinfo  {journal} {\textit{CLEO: QELS\_Fundamental Science}
  (Optical Society of America, 2019),}\ ,\ \bibinfo {pages} {paper
  FF1A}}}\bibinfo {note} {DOI: 10.1364/CLEO\_QELS.2019.FF1A.3}\BibitemShut
  {NoStop}%
\bibitem [{\citenamefont {Reddy}\ \emph {et~al.}(2020)\citenamefont {Reddy},
  \citenamefont {Nerem}, \citenamefont {Nam}, \citenamefont {Mirin},\ and\
  \citenamefont {Verma}}]{reddy_superconducting_2020}%
  \BibitemOpen
  \bibfield  {author} {\bibinfo {author} {\bibfnamefont {D.~V.}\ \bibnamefont
  {Reddy}}, \bibinfo {author} {\bibfnamefont {R.~R.}\ \bibnamefont {Nerem}},
  \bibinfo {author} {\bibfnamefont {S.~W.}\ \bibnamefont {Nam}}, \bibinfo
  {author} {\bibfnamefont {R.~P.}\ \bibnamefont {Mirin}},\ and\ \bibinfo
  {author} {\bibfnamefont {V.~B.}\ \bibnamefont {Verma}},\ }\bibfield  {title}
  {\bibinfo {title} {Superconducting nanowire single-photon detectors with 98\%
  system detection efficiency at 1550 nm},\ }\href@noop {} {\bibfield
  {journal} {\bibinfo  {journal} {Optica}\ }\textbf {\bibinfo {volume} {7}},\
  \bibinfo {pages} {1649} (\bibinfo {year} {2020})}\BibitemShut {NoStop}%
\bibitem [{\citenamefont {Chang}\ \emph {et~al.}(2021)\citenamefont {Chang},
  \citenamefont {Los}, \citenamefont {Tenorio-Pearl}, \citenamefont {Noordzij},
  \citenamefont {Gourgues}, \citenamefont {Guardiani}, \citenamefont {Zichi},
  \citenamefont {Pereira}, \citenamefont {Urbach}, \citenamefont {Zwiller},
  \citenamefont {Dorenbos},\ and\ \citenamefont
  {Esmaeil~Zadeh}}]{chang_detecting_2020}%
  \BibitemOpen
  \bibfield  {author} {\bibinfo {author} {\bibfnamefont {J.}~\bibnamefont
  {Chang}}, \bibinfo {author} {\bibfnamefont {J.~W.~N.}\ \bibnamefont {Los}},
  \bibinfo {author} {\bibfnamefont {J.~O.}\ \bibnamefont {Tenorio-Pearl}},
  \bibinfo {author} {\bibfnamefont {N.}~\bibnamefont {Noordzij}}, \bibinfo
  {author} {\bibfnamefont {R.}~\bibnamefont {Gourgues}}, \bibinfo {author}
  {\bibfnamefont {A.}~\bibnamefont {Guardiani}}, \bibinfo {author}
  {\bibfnamefont {J.~R.}\ \bibnamefont {Zichi}}, \bibinfo {author}
  {\bibfnamefont {S.~F.}\ \bibnamefont {Pereira}}, \bibinfo {author}
  {\bibfnamefont {H.~P.}\ \bibnamefont {Urbach}}, \bibinfo {author}
  {\bibfnamefont {V.}~\bibnamefont {Zwiller}}, \bibinfo {author} {\bibfnamefont
  {S.~N.}\ \bibnamefont {Dorenbos}},\ and\ \bibinfo {author} {\bibfnamefont
  {I.}~\bibnamefont {Esmaeil~Zadeh}},\ }\bibfield  {title} {\bibinfo {title}
  {Detecting telecom single photons with $\left(99.5_{-2.07}^{+0.5}\right)\%$
  system detection efficiency and high time resolution},\ }\href@noop {}
  {\bibfield  {journal} {\bibinfo  {journal} {APL Photonics}\ }\textbf
  {\bibinfo {volume} {6}},\ \bibinfo {pages} {036114} (\bibinfo {year}
  {2021})}\BibitemShut {NoStop}%
\bibitem [{\citenamefont {Hu}\ \emph {et~al.}(2020)\citenamefont {Hu},
  \citenamefont {Li}, \citenamefont {You}, \citenamefont {Wang}, \citenamefont
  {Xiao}, \citenamefont {Huang}, \citenamefont {Yang}, \citenamefont {Zhang},
  \citenamefont {Wang},\ and\ \citenamefont {Xie}}]{hu_detecting_2020}%
  \BibitemOpen
  \bibfield  {author} {\bibinfo {author} {\bibfnamefont {P.}~\bibnamefont
  {Hu}}, \bibinfo {author} {\bibfnamefont {H.}~\bibnamefont {Li}}, \bibinfo
  {author} {\bibfnamefont {L.}~\bibnamefont {You}}, \bibinfo {author}
  {\bibfnamefont {H.}~\bibnamefont {Wang}}, \bibinfo {author} {\bibfnamefont
  {Y.}~\bibnamefont {Xiao}}, \bibinfo {author} {\bibfnamefont {J.}~\bibnamefont
  {Huang}}, \bibinfo {author} {\bibfnamefont {X.}~\bibnamefont {Yang}},
  \bibinfo {author} {\bibfnamefont {W.}~\bibnamefont {Zhang}}, \bibinfo
  {author} {\bibfnamefont {Z.}~\bibnamefont {Wang}},\ and\ \bibinfo {author}
  {\bibfnamefont {X.}~\bibnamefont {Xie}},\ }\bibfield  {title} {\bibinfo
  {title} {Detecting single infrared photons toward optimal system detection
  efficiency},\ }\href@noop {} {\bibfield  {journal} {\bibinfo  {journal} {Opt.
  Express}\ }\textbf {\bibinfo {volume} {28}},\ \bibinfo {pages} {36884}
  (\bibinfo {year} {2020})}\BibitemShut {NoStop}%
\bibitem [{\citenamefont {Hochberg}\ \emph {et~al.}(2019)\citenamefont
  {Hochberg}, \citenamefont {Charaev}, \citenamefont {Nam}, \citenamefont
  {Verma}, \citenamefont {Colangelo},\ and\ \citenamefont
  {Berggren}}]{hochberg_detecting_2019}%
  \BibitemOpen
  \bibfield  {author} {\bibinfo {author} {\bibfnamefont {Y.}~\bibnamefont
  {Hochberg}}, \bibinfo {author} {\bibfnamefont {I.}~\bibnamefont {Charaev}},
  \bibinfo {author} {\bibfnamefont {S.-W.}\ \bibnamefont {Nam}}, \bibinfo
  {author} {\bibfnamefont {V.}~\bibnamefont {Verma}}, \bibinfo {author}
  {\bibfnamefont {M.}~\bibnamefont {Colangelo}},\ and\ \bibinfo {author}
  {\bibfnamefont {K.~K.}\ \bibnamefont {Berggren}},\ }\bibfield  {title}
  {\bibinfo {title} {Detecting sub-gev dark matter with superconducting
  nanowires},\ }\href@noop {} {\bibfield  {journal} {\bibinfo  {journal} {Phys.
  Rev. Lett.}\ }\textbf {\bibinfo {volume} {123}},\ \bibinfo {pages} {151802}
  (\bibinfo {year} {2019})}\BibitemShut {NoStop}%
\bibitem [{\citenamefont {Zhao}\ \emph {et~al.}(2014)\citenamefont {Zhao},
  \citenamefont {Jia}, \citenamefont {Gu}, \citenamefont {Wan}, \citenamefont
  {Zhang}, \citenamefont {Xu}, \citenamefont {Kang}, \citenamefont {Chen},\
  and\ \citenamefont {Wu}}]{zhao_counting_nodate}%
  \BibitemOpen
  \bibfield  {author} {\bibinfo {author} {\bibfnamefont {Q.}~\bibnamefont
  {Zhao}}, \bibinfo {author} {\bibfnamefont {T.}~\bibnamefont {Jia}}, \bibinfo
  {author} {\bibfnamefont {M.}~\bibnamefont {Gu}}, \bibinfo {author}
  {\bibfnamefont {C.}~\bibnamefont {Wan}}, \bibinfo {author} {\bibfnamefont
  {L.}~\bibnamefont {Zhang}}, \bibinfo {author} {\bibfnamefont
  {W.}~\bibnamefont {Xu}}, \bibinfo {author} {\bibfnamefont {L.}~\bibnamefont
  {Kang}}, \bibinfo {author} {\bibfnamefont {J.}~\bibnamefont {Chen}},\ and\
  \bibinfo {author} {\bibfnamefont {P.}~\bibnamefont {Wu}},\ }\bibfield
  {title} {\bibinfo {title} {Counting rate enhancements in superconducting
  nanowire single-photon detectors with improved readout circuits},\
  }\href@noop {} {\bibfield  {journal} {\bibinfo  {journal} {Opt. Lett.}\
  }\textbf {\bibinfo {volume} {39}},\ \bibinfo {pages} {1869} (\bibinfo {year}
  {2014})}\BibitemShut {NoStop}%
\bibitem [{\citenamefont {M{\"u}nzberg}\ \emph {et~al.}(2018)\citenamefont
  {M{\"u}nzberg}, \citenamefont {Vetter}, \citenamefont {Beutel}, \citenamefont
  {Hartmann}, \citenamefont {Ferrari}, \citenamefont {Pernice},\ and\
  \citenamefont {Rockstuhl}}]{munzberg2018superconducting}%
  \BibitemOpen
  \bibfield  {author} {\bibinfo {author} {\bibfnamefont {J.}~\bibnamefont
  {M{\"u}nzberg}}, \bibinfo {author} {\bibfnamefont {A.}~\bibnamefont
  {Vetter}}, \bibinfo {author} {\bibfnamefont {F.}~\bibnamefont {Beutel}},
  \bibinfo {author} {\bibfnamefont {W.}~\bibnamefont {Hartmann}}, \bibinfo
  {author} {\bibfnamefont {S.}~\bibnamefont {Ferrari}}, \bibinfo {author}
  {\bibfnamefont {W.~H.}\ \bibnamefont {Pernice}},\ and\ \bibinfo {author}
  {\bibfnamefont {C.}~\bibnamefont {Rockstuhl}},\ }\bibfield  {title} {\bibinfo
  {title} {Superconducting nanowire single-photon detector implemented in a 2d
  photonic crystal cavity},\ }\href@noop {} {\bibfield  {journal} {\bibinfo
  {journal} {Optica}\ }\textbf {\bibinfo {volume} {5}},\ \bibinfo {pages} {658}
  (\bibinfo {year} {2018})}\BibitemShut {NoStop}%
\bibitem [{\citenamefont {Tao}\ \emph {et~al.}(2019)\citenamefont {Tao},
  \citenamefont {Chen}, \citenamefont {Chen}, \citenamefont {Wang},
  \citenamefont {Li}, \citenamefont {Tu}, \citenamefont {Jia}, \citenamefont
  {Zhao}, \citenamefont {Zhang}, \citenamefont {Kang},\ and\ \citenamefont
  {Wu}}]{tao_high_2019}%
  \BibitemOpen
  \bibfield  {author} {\bibinfo {author} {\bibfnamefont {X.}~\bibnamefont
  {Tao}}, \bibinfo {author} {\bibfnamefont {S.}~\bibnamefont {Chen}}, \bibinfo
  {author} {\bibfnamefont {Y.}~\bibnamefont {Chen}}, \bibinfo {author}
  {\bibfnamefont {L.}~\bibnamefont {Wang}}, \bibinfo {author} {\bibfnamefont
  {X.}~\bibnamefont {Li}}, \bibinfo {author} {\bibfnamefont {X.}~\bibnamefont
  {Tu}}, \bibinfo {author} {\bibfnamefont {X.}~\bibnamefont {Jia}}, \bibinfo
  {author} {\bibfnamefont {Q.}~\bibnamefont {Zhao}}, \bibinfo {author}
  {\bibfnamefont {L.}~\bibnamefont {Zhang}}, \bibinfo {author} {\bibfnamefont
  {L.}~\bibnamefont {Kang}},\ and\ \bibinfo {author} {\bibfnamefont
  {P.}~\bibnamefont {Wu}},\ }\bibfield  {title} {\bibinfo {title} {A high speed
  and high efficiency superconducting photon number resolving detector},\
  }\href@noop {} {\bibfield  {journal} {\bibinfo  {journal} {Supercond. Sci.
  Technol.}\ }\textbf {\bibinfo {volume} {32}},\ \bibinfo {pages} {064002}
  (\bibinfo {year} {2019})}\BibitemShut {NoStop}%
\bibitem [{\citenamefont {Zhang}\ \emph {et~al.}(2019)\citenamefont {Zhang},
  \citenamefont {Huang}, \citenamefont {Zhang}, \citenamefont {You},
  \citenamefont {Lv}, \citenamefont {Zhang}, \citenamefont {Li}, \citenamefont
  {Wang},\ and\ \citenamefont {Xie}}]{zhang_16-pixel_2019}%
  \BibitemOpen
  \bibfield  {author} {\bibinfo {author} {\bibfnamefont {W.}~\bibnamefont
  {Zhang}}, \bibinfo {author} {\bibfnamefont {J.}~\bibnamefont {Huang}},
  \bibinfo {author} {\bibfnamefont {C.}~\bibnamefont {Zhang}}, \bibinfo
  {author} {\bibfnamefont {L.}~\bibnamefont {You}}, \bibinfo {author}
  {\bibfnamefont {C.}~\bibnamefont {Lv}}, \bibinfo {author} {\bibfnamefont
  {L.}~\bibnamefont {Zhang}}, \bibinfo {author} {\bibfnamefont
  {H.}~\bibnamefont {Li}}, \bibinfo {author} {\bibfnamefont {Z.}~\bibnamefont
  {Wang}},\ and\ \bibinfo {author} {\bibfnamefont {X.}~\bibnamefont {Xie}},\
  }\bibfield  {title} {\bibinfo {title} {A 16-pixel interleaved superconducting
  nanowire single-photon detector array with a maximum count rate exceeding
  1.5~$\textrm{GHz}$},\ }\href@noop {} {\bibfield  {journal} {\bibinfo
  {journal} {IEEE Trans. Appl. Supercond.}\ }\textbf {\bibinfo {volume} {29}},\
  \bibinfo {pages} {1} (\bibinfo {year} {2019})}\BibitemShut {NoStop}%
\bibitem [{\citenamefont {{Korzh}}\ \emph {et~al.}(2020)\citenamefont
  {{Korzh}}, \citenamefont {{Zhao}}, \citenamefont {{Allmaras}}, \citenamefont
  {{Frasca}}, \citenamefont {{Autry}}, \citenamefont {{Bersin}}, \citenamefont
  {{Beyer}}, \citenamefont {{Briggs}}, \citenamefont {{Bumble}}, \citenamefont
  {{Colangelo}}, \citenamefont {{Crouch}}, \citenamefont {{Dane}},
  \citenamefont {{Gerrits}}, \citenamefont {{Lita}}, \citenamefont {{Marsili}},
  \citenamefont {{Moody}}, \citenamefont {{Peña}}, \citenamefont {{Ramirez}},
  \citenamefont {{Rezac}}, \citenamefont {{Sinclair}}, \citenamefont
  {{Stevens}}, \citenamefont {{Velasco}}, \citenamefont {{Verma}},
  \citenamefont {{Wollman}}, \citenamefont {{Xie}}, \citenamefont {{Zhu}},
  \citenamefont {{Hale}}, \citenamefont {{Spiropulu}}, \citenamefont
  {{Silverman}}, \citenamefont {{Mirin}}, \citenamefont {{Nam}}, \citenamefont
  {{Kozorezov}}, \citenamefont {{Shaw}},\ and\ \citenamefont
  {{Berggren}}}]{korzh_demonstration_2020}%
  \BibitemOpen
  \bibfield  {author} {\bibinfo {author} {\bibfnamefont {B.}~\bibnamefont
  {{Korzh}}}, \bibinfo {author} {\bibfnamefont {Q.-Y.}\ \bibnamefont {{Zhao}}},
  \bibinfo {author} {\bibfnamefont {J.~P.}\ \bibnamefont {{Allmaras}}},
  \bibinfo {author} {\bibfnamefont {S.}~\bibnamefont {{Frasca}}}, \bibinfo
  {author} {\bibfnamefont {T.~M.}\ \bibnamefont {{Autry}}}, \bibinfo {author}
  {\bibfnamefont {E.~A.}\ \bibnamefont {{Bersin}}}, \bibinfo {author}
  {\bibfnamefont {A.~D.}\ \bibnamefont {{Beyer}}}, \bibinfo {author}
  {\bibfnamefont {R.~M.}\ \bibnamefont {{Briggs}}}, \bibinfo {author}
  {\bibfnamefont {B.}~\bibnamefont {{Bumble}}}, \bibinfo {author}
  {\bibfnamefont {M.}~\bibnamefont {{Colangelo}}}, \bibinfo {author}
  {\bibfnamefont {G.~M.}\ \bibnamefont {{Crouch}}}, \bibinfo {author}
  {\bibfnamefont {A.~E.}\ \bibnamefont {{Dane}}}, \bibinfo {author}
  {\bibfnamefont {T.}~\bibnamefont {{Gerrits}}}, \bibinfo {author}
  {\bibfnamefont {A.~E.}\ \bibnamefont {{Lita}}}, \bibinfo {author}
  {\bibfnamefont {F.}~\bibnamefont {{Marsili}}}, \bibinfo {author}
  {\bibfnamefont {G.}~\bibnamefont {{Moody}}}, \bibinfo {author} {\bibfnamefont
  {C.}~\bibnamefont {{Peña}}}, \bibinfo {author} {\bibfnamefont
  {E.}~\bibnamefont {{Ramirez}}}, \bibinfo {author} {\bibfnamefont {J.~D.}\
  \bibnamefont {{Rezac}}}, \bibinfo {author} {\bibfnamefont {N.}~\bibnamefont
  {{Sinclair}}}, \bibinfo {author} {\bibfnamefont {M.~J.}\ \bibnamefont
  {{Stevens}}}, \bibinfo {author} {\bibfnamefont {A.~E.}\ \bibnamefont
  {{Velasco}}}, \bibinfo {author} {\bibfnamefont {V.~B.}\ \bibnamefont
  {{Verma}}}, \bibinfo {author} {\bibfnamefont {E.~E.}\ \bibnamefont
  {{Wollman}}}, \bibinfo {author} {\bibfnamefont {S.}~\bibnamefont {{Xie}}},
  \bibinfo {author} {\bibfnamefont {D.}~\bibnamefont {{Zhu}}}, \bibinfo
  {author} {\bibfnamefont {P.~D.}\ \bibnamefont {{Hale}}}, \bibinfo {author}
  {\bibfnamefont {M.}~\bibnamefont {{Spiropulu}}}, \bibinfo {author}
  {\bibfnamefont {K.~L.}\ \bibnamefont {{Silverman}}}, \bibinfo {author}
  {\bibfnamefont {R.~P.}\ \bibnamefont {{Mirin}}}, \bibinfo {author}
  {\bibfnamefont {S.~W.}\ \bibnamefont {{Nam}}}, \bibinfo {author}
  {\bibfnamefont {A.~G.}\ \bibnamefont {{Kozorezov}}}, \bibinfo {author}
  {\bibfnamefont {M.~D.}\ \bibnamefont {{Shaw}}},\ and\ \bibinfo {author}
  {\bibfnamefont {K.~K.}\ \bibnamefont {{Berggren}}},\ }\bibfield  {title}
  {\bibinfo {title} {Demonstration of sub-3 ps temporal resolution with a
  superconducting nanowire single-photon detector},\ }\href@noop {} {\bibfield
  {journal} {\bibinfo  {journal} {Nat. Photonics}\ }\textbf {\bibinfo {volume}
  {14}},\ \bibinfo {pages} {250} (\bibinfo {year} {2020})}\BibitemShut
  {NoStop}%
\bibitem [{\citenamefont {Esmaeil~Zadeh}\ \emph {et~al.}(2020)\citenamefont
  {Esmaeil~Zadeh}, \citenamefont {Los}, \citenamefont {Gourgues}, \citenamefont
  {Chang}, \citenamefont {Elshaari}, \citenamefont {Zichi}, \citenamefont {van
  Staaden}, \citenamefont {Swens}, \citenamefont {Kalhor}, \citenamefont
  {Guardiani}, \citenamefont {Meng}, \citenamefont {Zou}, \citenamefont
  {Dobrovolskiy}, \citenamefont {Fognini}, \citenamefont {Schaart},
  \citenamefont {Dalacu}, \citenamefont {Poole}, \citenamefont {Reimer},
  \citenamefont {Hu}, \citenamefont {Pereira}, \citenamefont {Zwiller},\ and\
  \citenamefont {Dorenbos}}]{esmaeil_zadeh_efficient_2020}%
  \BibitemOpen
  \bibfield  {author} {\bibinfo {author} {\bibfnamefont {I.}~\bibnamefont
  {Esmaeil~Zadeh}}, \bibinfo {author} {\bibfnamefont {J.~W.~N.}\ \bibnamefont
  {Los}}, \bibinfo {author} {\bibfnamefont {R.~B.~M.}\ \bibnamefont
  {Gourgues}}, \bibinfo {author} {\bibfnamefont {J.}~\bibnamefont {Chang}},
  \bibinfo {author} {\bibfnamefont {A.~W.}\ \bibnamefont {Elshaari}}, \bibinfo
  {author} {\bibfnamefont {J.~R.}\ \bibnamefont {Zichi}}, \bibinfo {author}
  {\bibfnamefont {Y.~J.}\ \bibnamefont {van Staaden}}, \bibinfo {author}
  {\bibfnamefont {J.~P.~E.}\ \bibnamefont {Swens}}, \bibinfo {author}
  {\bibfnamefont {N.}~\bibnamefont {Kalhor}}, \bibinfo {author} {\bibfnamefont
  {A.}~\bibnamefont {Guardiani}}, \bibinfo {author} {\bibfnamefont
  {Y.}~\bibnamefont {Meng}}, \bibinfo {author} {\bibfnamefont {K.}~\bibnamefont
  {Zou}}, \bibinfo {author} {\bibfnamefont {S.}~\bibnamefont {Dobrovolskiy}},
  \bibinfo {author} {\bibfnamefont {A.~W.}\ \bibnamefont {Fognini}}, \bibinfo
  {author} {\bibfnamefont {D.~R.}\ \bibnamefont {Schaart}}, \bibinfo {author}
  {\bibfnamefont {D.}~\bibnamefont {Dalacu}}, \bibinfo {author} {\bibfnamefont
  {P.~J.}\ \bibnamefont {Poole}}, \bibinfo {author} {\bibfnamefont {M.~E.}\
  \bibnamefont {Reimer}}, \bibinfo {author} {\bibfnamefont {X.}~\bibnamefont
  {Hu}}, \bibinfo {author} {\bibfnamefont {S.~F.}\ \bibnamefont {Pereira}},
  \bibinfo {author} {\bibfnamefont {V.}~\bibnamefont {Zwiller}},\ and\ \bibinfo
  {author} {\bibfnamefont {S.~N.}\ \bibnamefont {Dorenbos}},\ }\bibfield
  {title} {\bibinfo {title} {Efficient single-photon detection with 7.7 ps time
  resolution for photon correlation measurements},\ }\href@noop {} {\bibfield
  {journal} {\bibinfo  {journal} {ACS Photonics}\ }\textbf {\bibinfo {volume}
  {7}},\ \bibinfo {pages} {1780} (\bibinfo {year} {2020})}\BibitemShut
  {NoStop}%
\bibitem [{\citenamefont {Marsili}\ \emph {et~al.}(2012)\citenamefont
  {Marsili}, \citenamefont {Bellei}, \citenamefont {Najafi}, \citenamefont
  {Dane}, \citenamefont {Dauler}, \citenamefont {Molnar},\ and\ \citenamefont
  {Berggren}}]{marsili_efficient_2012}%
  \BibitemOpen
  \bibfield  {author} {\bibinfo {author} {\bibfnamefont {F.}~\bibnamefont
  {Marsili}}, \bibinfo {author} {\bibfnamefont {F.}~\bibnamefont {Bellei}},
  \bibinfo {author} {\bibfnamefont {F.}~\bibnamefont {Najafi}}, \bibinfo
  {author} {\bibfnamefont {A.~E.}\ \bibnamefont {Dane}}, \bibinfo {author}
  {\bibfnamefont {E.~A.}\ \bibnamefont {Dauler}}, \bibinfo {author}
  {\bibfnamefont {R.~J.}\ \bibnamefont {Molnar}},\ and\ \bibinfo {author}
  {\bibfnamefont {K.~K.}\ \bibnamefont {Berggren}},\ }\bibfield  {title}
  {\bibinfo {title} {Efficient single photon detection from 500 nm to 5~\textmu
  m wavelength},\ }\href@noop {} {\bibfield  {journal} {\bibinfo  {journal}
  {Nano Lett.}\ }\textbf {\bibinfo {volume} {12}},\ \bibinfo {pages} {4799}
  (\bibinfo {year} {2012})}\BibitemShut {NoStop}%
\bibitem [{\citenamefont {{Verma}}\ \emph {et~al.}(2019)\citenamefont
  {{Verma}}, \citenamefont {{Lita}}, \citenamefont {{Korzh}}, \citenamefont
  {{Wollman}}, \citenamefont {{Shaw}}, \citenamefont {{Mirin}},\ and\
  \citenamefont {{Nam}}}]{verma_towards_2019}%
  \BibitemOpen
  \bibfield  {author} {\bibinfo {author} {\bibfnamefont {V.~B.}\ \bibnamefont
  {{Verma}}}, \bibinfo {author} {\bibfnamefont {A.~E.}\ \bibnamefont {{Lita}}},
  \bibinfo {author} {\bibfnamefont {B.~A.}\ \bibnamefont {{Korzh}}}, \bibinfo
  {author} {\bibfnamefont {E.}~\bibnamefont {{Wollman}}}, \bibinfo {author}
  {\bibfnamefont {M.}~\bibnamefont {{Shaw}}}, \bibinfo {author} {\bibfnamefont
  {R.~P.}\ \bibnamefont {{Mirin}}},\ and\ \bibinfo {author} {\bibfnamefont
  {S.-W.}\ \bibnamefont {{Nam}}},\ }\bibfield  {title} {\bibinfo {title}
  {Towards single-photon spectroscopy in the mid-infrared using superconducting
  nanowire single-photon detectors},\ }\href@noop {} {\bibfield  {journal}
  {\bibinfo  {journal} {\textit{Advanced Photon Counting Techniques XIII}
  (International Society for Optics and Photonics 2019),}\ ,\ \bibinfo {pages}
  {paper 109780N}} (\bibinfo {year} {2019})},\ \bibinfo {note} {dOI:
  10.1117/12.2519474}\BibitemShut {NoStop}%
\bibitem [{\citenamefont {Chen}\ \emph {et~al.}(2021)\citenamefont {Chen},
  \citenamefont {Ge}, \citenamefont {Zhang}, \citenamefont {Li}, \citenamefont
  {Zhang}, \citenamefont {Dai}, \citenamefont {Fei}, \citenamefont {Wang},
  \citenamefont {Jia}, \citenamefont {Zhao}, \citenamefont {Tu}, \citenamefont
  {Kang}, \citenamefont {Chen},\ and\ \citenamefont
  {Wu}}]{chen_mid-infrared_2020}%
  \BibitemOpen
  \bibfield  {author} {\bibinfo {author} {\bibfnamefont {Q.}~\bibnamefont
  {Chen}}, \bibinfo {author} {\bibfnamefont {R.}~\bibnamefont {Ge}}, \bibinfo
  {author} {\bibfnamefont {L.}~\bibnamefont {Zhang}}, \bibinfo {author}
  {\bibfnamefont {F.}~\bibnamefont {Li}}, \bibinfo {author} {\bibfnamefont
  {B.}~\bibnamefont {Zhang}}, \bibinfo {author} {\bibfnamefont
  {Y.}~\bibnamefont {Dai}}, \bibinfo {author} {\bibfnamefont {Y.}~\bibnamefont
  {Fei}}, \bibinfo {author} {\bibfnamefont {X.}~\bibnamefont {Wang}}, \bibinfo
  {author} {\bibfnamefont {X.}~\bibnamefont {Jia}}, \bibinfo {author}
  {\bibfnamefont {Q.}~\bibnamefont {Zhao}}, \bibinfo {author} {\bibfnamefont
  {X.}~\bibnamefont {Tu}}, \bibinfo {author} {\bibfnamefont {L.}~\bibnamefont
  {Kang}}, \bibinfo {author} {\bibfnamefont {J.}~\bibnamefont {Chen}},\ and\
  \bibinfo {author} {\bibfnamefont {P.}~\bibnamefont {Wu}},\ }\bibfield
  {title} {\bibinfo {title} {Mid-infrared single photon detector with
  superconductor $\rm{Mo_{0.8}Si_{0.2}}$ nanowire},\ }\href@noop {} {\bibfield
  {journal} {\bibinfo  {journal} {Sci. Bull.}\ }\textbf {\bibinfo {volume}
  {66}},\ \bibinfo {pages} {965} (\bibinfo {year} {2021})}\BibitemShut
  {NoStop}%
\bibitem [{\citenamefont {Zhang}\ \emph {et~al.}(2016)\citenamefont {Zhang},
  \citenamefont {Wang},\ and\ \citenamefont
  {Schilling}}]{zhang_superconducting_2016}%
  \BibitemOpen
  \bibfield  {author} {\bibinfo {author} {\bibfnamefont {X.}~\bibnamefont
  {Zhang}}, \bibinfo {author} {\bibfnamefont {Q.}~\bibnamefont {Wang}},\ and\
  \bibinfo {author} {\bibfnamefont {A.}~\bibnamefont {Schilling}},\ }\bibfield
  {title} {\bibinfo {title} {Superconducting single $\textrm{X}$-ray photon
  detector based on $\textrm{W}_{0.8}\textrm{Si}_{0.2}$},\ }\href@noop {}
  {\bibfield  {journal} {\bibinfo  {journal} {AIP Adv.}\ }\textbf {\bibinfo
  {volume} {6}},\ \bibinfo {pages} {115104} (\bibinfo {year}
  {2016})}\BibitemShut {NoStop}%
\bibitem [{\citenamefont {Gol’tsman}\ \emph {et~al.}(2001)\citenamefont
  {Gol’tsman}, \citenamefont {Okunev}, \citenamefont {Chulkova},
  \citenamefont {Lipatov}, \citenamefont {Semenov}, \citenamefont {Smirnov},
  \citenamefont {Voronov}, \citenamefont {Dzardanov}, \citenamefont
  {Williams},\ and\ \citenamefont {Sobolewski}}]{goltsman_picosecond_2001}%
  \BibitemOpen
  \bibfield  {author} {\bibinfo {author} {\bibfnamefont {G.~N.}\ \bibnamefont
  {Gol’tsman}}, \bibinfo {author} {\bibfnamefont {O.}~\bibnamefont {Okunev}},
  \bibinfo {author} {\bibfnamefont {G.}~\bibnamefont {Chulkova}}, \bibinfo
  {author} {\bibfnamefont {A.}~\bibnamefont {Lipatov}}, \bibinfo {author}
  {\bibfnamefont {A.}~\bibnamefont {Semenov}}, \bibinfo {author} {\bibfnamefont
  {K.}~\bibnamefont {Smirnov}}, \bibinfo {author} {\bibfnamefont
  {B.}~\bibnamefont {Voronov}}, \bibinfo {author} {\bibfnamefont
  {A.}~\bibnamefont {Dzardanov}}, \bibinfo {author} {\bibfnamefont
  {C.}~\bibnamefont {Williams}},\ and\ \bibinfo {author} {\bibfnamefont
  {R.}~\bibnamefont {Sobolewski}},\ }\bibfield  {title} {\bibinfo {title}
  {Picosecond superconducting single-photon optical detector},\ }\href@noop {}
  {\bibfield  {journal} {\bibinfo  {journal} {Appl. Phys. Lett.}\ }\textbf
  {\bibinfo {volume} {79}},\ \bibinfo {pages} {705} (\bibinfo {year}
  {2001})}\BibitemShut {NoStop}%
\bibitem [{\citenamefont {Hadfield}(2009)}]{hadfield_single-photon_2009}%
  \BibitemOpen
  \bibfield  {author} {\bibinfo {author} {\bibfnamefont {R.~H.}\ \bibnamefont
  {Hadfield}},\ }\bibfield  {title} {\bibinfo {title} {single-photon detectors
  for optical quantum information applications},\ }\href@noop {} {\bibfield
  {journal} {\bibinfo  {journal} {Nat. Photonics}\ }\textbf {\bibinfo {volume}
  {3}},\ \bibinfo {pages} {696} (\bibinfo {year} {2009})}\BibitemShut {NoStop}%
\bibitem [{\citenamefont {Taylor}\ \emph {et~al.}(2019)\citenamefont {Taylor},
  \citenamefont {Morozov}, \citenamefont {Gemmell}, \citenamefont
  {Erotokritou}, \citenamefont {Miki}, \citenamefont {Terai},\ and\
  \citenamefont {Hadfield}}]{taylor2019photon}%
  \BibitemOpen
  \bibfield  {author} {\bibinfo {author} {\bibfnamefont {G.~G.}\ \bibnamefont
  {Taylor}}, \bibinfo {author} {\bibfnamefont {D.}~\bibnamefont {Morozov}},
  \bibinfo {author} {\bibfnamefont {N.~R.}\ \bibnamefont {Gemmell}}, \bibinfo
  {author} {\bibfnamefont {K.}~\bibnamefont {Erotokritou}}, \bibinfo {author}
  {\bibfnamefont {S.}~\bibnamefont {Miki}}, \bibinfo {author} {\bibfnamefont
  {H.}~\bibnamefont {Terai}},\ and\ \bibinfo {author} {\bibfnamefont {R.~H.}\
  \bibnamefont {Hadfield}},\ }\bibfield  {title} {\bibinfo {title} {Photon
  counting lidar at 2.3 \textmu m wavelength with superconducting nanowires},\
  }\href@noop {} {\bibfield  {journal} {\bibinfo  {journal} {Opt. Express}\
  }\textbf {\bibinfo {volume} {27}},\ \bibinfo {pages} {38147} (\bibinfo {year}
  {2019})}\BibitemShut {NoStop}%
\bibitem [{\citenamefont {Gemmell}\ \emph {et~al.}(2013)\citenamefont
  {Gemmell}, \citenamefont {McCarthy}, \citenamefont {Liu}, \citenamefont
  {Tanner}, \citenamefont {Dorenbos}, \citenamefont {Zwiller}, \citenamefont
  {Patterson}, \citenamefont {Buller}, \citenamefont {Wilson},\ and\
  \citenamefont {Hadfield}}]{gemmell2013singlet}%
  \BibitemOpen
  \bibfield  {author} {\bibinfo {author} {\bibfnamefont {N.~R.}\ \bibnamefont
  {Gemmell}}, \bibinfo {author} {\bibfnamefont {A.}~\bibnamefont {McCarthy}},
  \bibinfo {author} {\bibfnamefont {B.}~\bibnamefont {Liu}}, \bibinfo {author}
  {\bibfnamefont {M.~G.}\ \bibnamefont {Tanner}}, \bibinfo {author}
  {\bibfnamefont {S.~D.}\ \bibnamefont {Dorenbos}}, \bibinfo {author}
  {\bibfnamefont {V.}~\bibnamefont {Zwiller}}, \bibinfo {author} {\bibfnamefont
  {M.~S.}\ \bibnamefont {Patterson}}, \bibinfo {author} {\bibfnamefont {G.~S.}\
  \bibnamefont {Buller}}, \bibinfo {author} {\bibfnamefont {B.~C.}\
  \bibnamefont {Wilson}},\ and\ \bibinfo {author} {\bibfnamefont {R.~H.}\
  \bibnamefont {Hadfield}},\ }\bibfield  {title} {\bibinfo {title} {Singlet
  oxygen luminescence detection with a fiber-coupled superconducting nanowire
  single-photon detector},\ }\href@noop {} {\bibfield  {journal} {\bibinfo
  {journal} {Opt. Express}\ }\textbf {\bibinfo {volume} {21}},\ \bibinfo
  {pages} {5005} (\bibinfo {year} {2013})}\BibitemShut {NoStop}%
\bibitem [{\citenamefont {Yin}\ \emph {et~al.}(2016)\citenamefont {Yin},
  \citenamefont {Chen}, \citenamefont {Yu}, \citenamefont {Liu}, \citenamefont
  {You}, \citenamefont {Zhou}, \citenamefont {Chen}, \citenamefont {Mao},
  \citenamefont {Huang}, \citenamefont {Zhang}, \citenamefont {Chen},
  \citenamefont {Li}, \citenamefont {Nolan}, \citenamefont {Zhou},
  \citenamefont {Jiang}, \citenamefont {Wang}, \citenamefont {Zhang},
  \citenamefont {Wang},\ and\ \citenamefont {Pan}}]{yin2016measurement}%
  \BibitemOpen
  \bibfield  {author} {\bibinfo {author} {\bibfnamefont {H.~L.}\ \bibnamefont
  {Yin}}, \bibinfo {author} {\bibfnamefont {T.-Y.}\ \bibnamefont {Chen}},
  \bibinfo {author} {\bibfnamefont {Z.-W.}\ \bibnamefont {Yu}}, \bibinfo
  {author} {\bibfnamefont {H.}~\bibnamefont {Liu}}, \bibinfo {author}
  {\bibfnamefont {L.-X.}\ \bibnamefont {You}}, \bibinfo {author} {\bibfnamefont
  {Y.-H.}\ \bibnamefont {Zhou}}, \bibinfo {author} {\bibfnamefont {S.-J.}\
  \bibnamefont {Chen}}, \bibinfo {author} {\bibfnamefont {Y.}~\bibnamefont
  {Mao}}, \bibinfo {author} {\bibfnamefont {M.-Q.}\ \bibnamefont {Huang}},
  \bibinfo {author} {\bibfnamefont {W.-J.}\ \bibnamefont {Zhang}}, \bibinfo
  {author} {\bibfnamefont {H.}~\bibnamefont {Chen}}, \bibinfo {author}
  {\bibfnamefont {M.-J.}\ \bibnamefont {Li}}, \bibinfo {author} {\bibfnamefont
  {D.}~\bibnamefont {Nolan}}, \bibinfo {author} {\bibfnamefont
  {F.}~\bibnamefont {Zhou}}, \bibinfo {author} {\bibfnamefont {X.}~\bibnamefont
  {Jiang}}, \bibinfo {author} {\bibfnamefont {Z.}~\bibnamefont {Wang}},
  \bibinfo {author} {\bibfnamefont {Q.}~\bibnamefont {Zhang}}, \bibinfo
  {author} {\bibfnamefont {X.-B.}\ \bibnamefont {Wang}},\ and\ \bibinfo
  {author} {\bibfnamefont {J.-W.}\ \bibnamefont {Pan}},\ }\bibfield  {title}
  {\bibinfo {title} {Measurement-device-independent quantum key distribution
  over a 404~km optical fiber},\ }\href@noop {} {\bibfield  {journal} {\bibinfo
   {journal} {Phys. Rev. Lett.}\ }\textbf {\bibinfo {volume} {117}},\ \bibinfo
  {pages} {190501} (\bibinfo {year} {2016})}\BibitemShut {NoStop}%
\bibitem [{\citenamefont {Wang}\ \emph {et~al.}(2019)\citenamefont {Wang},
  \citenamefont {Qin}, \citenamefont {Ding}, \citenamefont {Chen},
  \citenamefont {Chen}, \citenamefont {You}, \citenamefont {He}, \citenamefont
  {Jiang}, \citenamefont {You}, \citenamefont {Wang}, \citenamefont
  {Schneider}, \citenamefont {Renema}, \citenamefont {Höfling}, \citenamefont
  {Lu},\ and\ \citenamefont {Pan}}]{wang2019boson}%
  \BibitemOpen
  \bibfield  {author} {\bibinfo {author} {\bibfnamefont {H.}~\bibnamefont
  {Wang}}, \bibinfo {author} {\bibfnamefont {J.}~\bibnamefont {Qin}}, \bibinfo
  {author} {\bibfnamefont {X.}~\bibnamefont {Ding}}, \bibinfo {author}
  {\bibfnamefont {M.-C.}\ \bibnamefont {Chen}}, \bibinfo {author}
  {\bibfnamefont {S.}~\bibnamefont {Chen}}, \bibinfo {author} {\bibfnamefont
  {X.}~\bibnamefont {You}}, \bibinfo {author} {\bibfnamefont {Y.-M.}\
  \bibnamefont {He}}, \bibinfo {author} {\bibfnamefont {X.}~\bibnamefont
  {Jiang}}, \bibinfo {author} {\bibfnamefont {L.}~\bibnamefont {You}}, \bibinfo
  {author} {\bibfnamefont {Z.}~\bibnamefont {Wang}}, \bibinfo {author}
  {\bibfnamefont {C.}~\bibnamefont {Schneider}}, \bibinfo {author}
  {\bibfnamefont {J.~J.}\ \bibnamefont {Renema}}, \bibinfo {author}
  {\bibfnamefont {S.}~\bibnamefont {Höfling}}, \bibinfo {author}
  {\bibfnamefont {C.-Y.}\ \bibnamefont {Lu}},\ and\ \bibinfo {author}
  {\bibfnamefont {J.-W.}\ \bibnamefont {Pan}},\ }\bibfield  {title} {\bibinfo
  {title} {Boson sampling with 20 input photons and a 60-mode interferometer in
  a 10$^{14}$-dimensional hilbert space},\ }\href@noop {} {\bibfield  {journal}
  {\bibinfo  {journal} {Phys. Rev. Lett.}\ }\textbf {\bibinfo {volume} {123}},\
  \bibinfo {pages} {250503} (\bibinfo {year} {2019})}\BibitemShut {NoStop}%
\bibitem [{\citenamefont {{Zhong}}\ \emph {et~al.}(2020)\citenamefont
  {{Zhong}}, \citenamefont {{Wang}}, \citenamefont {{Deng}}, \citenamefont
  {{Chen}}, \citenamefont {{Peng}}, \citenamefont {{Luo}}, \citenamefont
  {{Qin}}, \citenamefont {{Wu}}, \citenamefont {{Ding}}, \citenamefont {{Hu}},
  \citenamefont {{Hu}}, \citenamefont {{Yang}}, \citenamefont {{Zhang}},
  \citenamefont {{Li}}, \citenamefont {{Li}}, \citenamefont {{Jiang}},
  \citenamefont {{Gan}}, \citenamefont {{Yang}}, \citenamefont {{You}},
  \citenamefont {{Wang}}, \citenamefont {{Li}}, \citenamefont {{Liu}},
  \citenamefont {{Lu}},\ and\ \citenamefont {{Pan}}}]{quantum2020}%
  \BibitemOpen
  \bibfield  {author} {\bibinfo {author} {\bibfnamefont {H.~S.}\ \bibnamefont
  {{Zhong}}}, \bibinfo {author} {\bibfnamefont {H.}~\bibnamefont {{Wang}}},
  \bibinfo {author} {\bibfnamefont {Y.-H.}\ \bibnamefont {{Deng}}}, \bibinfo
  {author} {\bibfnamefont {M.-C.}\ \bibnamefont {{Chen}}}, \bibinfo {author}
  {\bibfnamefont {L.-C.}\ \bibnamefont {{Peng}}}, \bibinfo {author}
  {\bibfnamefont {Y.-H.}\ \bibnamefont {{Luo}}}, \bibinfo {author}
  {\bibfnamefont {J.}~\bibnamefont {{Qin}}}, \bibinfo {author} {\bibfnamefont
  {D.}~\bibnamefont {{Wu}}}, \bibinfo {author} {\bibfnamefont {X.}~\bibnamefont
  {{Ding}}}, \bibinfo {author} {\bibfnamefont {Y.}~\bibnamefont {{Hu}}},
  \bibinfo {author} {\bibfnamefont {P.}~\bibnamefont {{Hu}}}, \bibinfo {author}
  {\bibfnamefont {X.-Y.}\ \bibnamefont {{Yang}}}, \bibinfo {author}
  {\bibfnamefont {W.-J.}\ \bibnamefont {{Zhang}}}, \bibinfo {author}
  {\bibfnamefont {H.}~\bibnamefont {{Li}}}, \bibinfo {author} {\bibfnamefont
  {Y.}~\bibnamefont {{Li}}}, \bibinfo {author} {\bibfnamefont {X.}~\bibnamefont
  {{Jiang}}}, \bibinfo {author} {\bibfnamefont {L.}~\bibnamefont {{Gan}}},
  \bibinfo {author} {\bibfnamefont {G.}~\bibnamefont {{Yang}}}, \bibinfo
  {author} {\bibfnamefont {L.}~\bibnamefont {{You}}}, \bibinfo {author}
  {\bibfnamefont {Z.}~\bibnamefont {{Wang}}}, \bibinfo {author} {\bibfnamefont
  {L.}~\bibnamefont {{Li}}}, \bibinfo {author} {\bibfnamefont {N.-L.}\
  \bibnamefont {{Liu}}}, \bibinfo {author} {\bibfnamefont {C.-Y.}\ \bibnamefont
  {{Lu}}},\ and\ \bibinfo {author} {\bibfnamefont {J.-W.}\ \bibnamefont
  {{Pan}}},\ }\bibfield  {title} {\bibinfo {title} {Quantum computational
  advantage using photons},\ }\href@noop {} {\bibfield  {journal} {\bibinfo
  {journal} {Science}\ }\textbf {\bibinfo {volume} {370}},\ \bibinfo {pages}
  {1460} (\bibinfo {year} {2020})}\BibitemShut {NoStop}%
\bibitem [{\citenamefont {Zou}\ \emph {et~al.}(2020{\natexlab{a}})\citenamefont
  {Zou}, \citenamefont {Meng}, \citenamefont {Wang},\ and\ \citenamefont
  {Hu}}]{zou2020superconductingPR}%
  \BibitemOpen
  \bibfield  {author} {\bibinfo {author} {\bibfnamefont {K.}~\bibnamefont
  {Zou}}, \bibinfo {author} {\bibfnamefont {Y.}~\bibnamefont {Meng}}, \bibinfo
  {author} {\bibfnamefont {Z.}~\bibnamefont {Wang}},\ and\ \bibinfo {author}
  {\bibfnamefont {X.}~\bibnamefont {Hu}},\ }\bibfield  {title} {\bibinfo
  {title} {Superconducting nanowire multi-photon detectors enabled by current
  reservoirs},\ }\href@noop {} {\bibfield  {journal} {\bibinfo  {journal}
  {Photonics Res.}\ }\textbf {\bibinfo {volume} {8}},\ \bibinfo {pages} {601}
  (\bibinfo {year} {2020}{\natexlab{a}})}\BibitemShut {NoStop}%
\bibitem [{\citenamefont {Zou}\ \emph {et~al.}(2020{\natexlab{b}})\citenamefont
  {Zou}, \citenamefont {Meng}, \citenamefont {Xu}, \citenamefont {Hu},
  \citenamefont {Wang},\ and\ \citenamefont
  {Hu}}]{zou2020superconductingPRApplied}%
  \BibitemOpen
  \bibfield  {author} {\bibinfo {author} {\bibfnamefont {K.}~\bibnamefont
  {Zou}}, \bibinfo {author} {\bibfnamefont {Y.}~\bibnamefont {Meng}}, \bibinfo
  {author} {\bibfnamefont {L.}~\bibnamefont {Xu}}, \bibinfo {author}
  {\bibfnamefont {N.}~\bibnamefont {Hu}}, \bibinfo {author} {\bibfnamefont
  {Z.}~\bibnamefont {Wang}},\ and\ \bibinfo {author} {\bibfnamefont
  {X.}~\bibnamefont {Hu}},\ }\bibfield  {title} {\bibinfo {title}
  {Superconducting nanowire photon-number-resolving detectors integrated with
  current reservoirs},\ }\href@noop {} {\bibfield  {journal} {\bibinfo
  {journal} {Phys. Rev. Appl.}\ }\textbf {\bibinfo {volume} {14}},\ \bibinfo
  {pages} {044029} (\bibinfo {year} {2020}{\natexlab{b}})}\BibitemShut
  {NoStop}%
\bibitem [{\citenamefont {Wei}\ \emph {et~al.}(2019)\citenamefont {Wei},
  \citenamefont {Zhang}, \citenamefont {Tang}, \citenamefont {You},\ and\
  \citenamefont {Xu}}]{wei2019implementation}%
  \BibitemOpen
  \bibfield  {author} {\bibinfo {author} {\bibfnamefont {K.}~\bibnamefont
  {Wei}}, \bibinfo {author} {\bibfnamefont {W.}~\bibnamefont {Zhang}}, \bibinfo
  {author} {\bibfnamefont {Y.~L.}\ \bibnamefont {Tang}}, \bibinfo {author}
  {\bibfnamefont {L.}~\bibnamefont {You}},\ and\ \bibinfo {author}
  {\bibfnamefont {F.}~\bibnamefont {Xu}},\ }\bibfield  {title} {\bibinfo
  {title} {Implementation security of quantum key distribution due to
  polarization-dependent efficiency mismatch},\ }\href@noop {} {\bibfield
  {journal} {\bibinfo  {journal} {Phys. Rev. A}\ }\textbf {\bibinfo {volume}
  {100}},\ \bibinfo {pages} {022325} (\bibinfo {year} {2019})}\BibitemShut
  {NoStop}%
\bibitem [{\citenamefont {Meng}\ \emph {et~al.}(2020)\citenamefont {Meng},
  \citenamefont {Zou}, \citenamefont {Hu}, \citenamefont {Lan}, \citenamefont
  {Xu}, \citenamefont {Zichi}, \citenamefont {Steinhauer}, \citenamefont
  {Zwiller},\ and\ \citenamefont {Hu}}]{meng_fractal_2020}%
  \BibitemOpen
  \bibfield  {author} {\bibinfo {author} {\bibfnamefont {Y.}~\bibnamefont
  {Meng}}, \bibinfo {author} {\bibfnamefont {K.}~\bibnamefont {Zou}}, \bibinfo
  {author} {\bibfnamefont {N.}~\bibnamefont {Hu}}, \bibinfo {author}
  {\bibfnamefont {X.}~\bibnamefont {Lan}}, \bibinfo {author} {\bibfnamefont
  {L.}~\bibnamefont {Xu}}, \bibinfo {author} {\bibfnamefont {J.}~\bibnamefont
  {Zichi}}, \bibinfo {author} {\bibfnamefont {S.}~\bibnamefont {Steinhauer}},
  \bibinfo {author} {\bibfnamefont {V.}~\bibnamefont {Zwiller}},\ and\ \bibinfo
  {author} {\bibfnamefont {X.}~\bibnamefont {Hu}},\ }\bibfield  {title}
  {\bibinfo {title} {Fractal superconducting nanowire avalanche photodetector
  at 1550 nm with 60\% system detection efficiency and 1.05 polarization
  sensitivity},\ }\href@noop {} {\bibfield  {journal} {\bibinfo  {journal}
  {Opt. Lett.}\ }\textbf {\bibinfo {volume} {45}},\ \bibinfo {pages} {471}
  (\bibinfo {year} {2020})}\BibitemShut {NoStop}%
\bibitem [{\citenamefont {Dorenbos}\ \emph {et~al.}(2008)\citenamefont
  {Dorenbos}, \citenamefont {Reiger}, \citenamefont {Akopian}, \citenamefont
  {Perinetti}, \citenamefont {Zwiller}, \citenamefont {Zijlstra},\ and\
  \citenamefont {Klapwijk}}]{dorenbos_superconducting_2008}%
  \BibitemOpen
  \bibfield  {author} {\bibinfo {author} {\bibfnamefont {S.~N.}\ \bibnamefont
  {Dorenbos}}, \bibinfo {author} {\bibfnamefont {E.~M.}\ \bibnamefont
  {Reiger}}, \bibinfo {author} {\bibfnamefont {N.}~\bibnamefont {Akopian}},
  \bibinfo {author} {\bibfnamefont {U.}~\bibnamefont {Perinetti}}, \bibinfo
  {author} {\bibfnamefont {V.}~\bibnamefont {Zwiller}}, \bibinfo {author}
  {\bibfnamefont {T.}~\bibnamefont {Zijlstra}},\ and\ \bibinfo {author}
  {\bibfnamefont {T.~M.}\ \bibnamefont {Klapwijk}},\ }\bibfield  {title}
  {\bibinfo {title} {Superconducting single photon detectors with minimized
  polarization dependence},\ }\href@noop {} {\bibfield  {journal} {\bibinfo
  {journal} {Appl. Phys. Lett.}\ }\textbf {\bibinfo {volume} {93}},\ \bibinfo
  {pages} {161102} (\bibinfo {year} {2008})}\BibitemShut {NoStop}%
\bibitem [{\citenamefont {Huang}\ \emph {et~al.}(2017)\citenamefont {Huang},
  \citenamefont {Zhang}, \citenamefont {You}, \citenamefont {Liu},
  \citenamefont {Guo}, \citenamefont {Wang}, \citenamefont {Zhang},
  \citenamefont {Yang}, \citenamefont {Li}, \citenamefont {Wang},\ and\
  \citenamefont {Xie}}]{huang_spiral_2017}%
  \BibitemOpen
  \bibfield  {author} {\bibinfo {author} {\bibfnamefont {J.}~\bibnamefont
  {Huang}}, \bibinfo {author} {\bibfnamefont {W.}~\bibnamefont {Zhang}},
  \bibinfo {author} {\bibfnamefont {L.}~\bibnamefont {You}}, \bibinfo {author}
  {\bibfnamefont {X.}~\bibnamefont {Liu}}, \bibinfo {author} {\bibfnamefont
  {Q.}~\bibnamefont {Guo}}, \bibinfo {author} {\bibfnamefont {Y.}~\bibnamefont
  {Wang}}, \bibinfo {author} {\bibfnamefont {L.}~\bibnamefont {Zhang}},
  \bibinfo {author} {\bibfnamefont {X.}~\bibnamefont {Yang}}, \bibinfo {author}
  {\bibfnamefont {H.}~\bibnamefont {Li}}, \bibinfo {author} {\bibfnamefont
  {Z.}~\bibnamefont {Wang}},\ and\ \bibinfo {author} {\bibfnamefont {X.~M.}\
  \bibnamefont {Xie}},\ }\bibfield  {title} {\bibinfo {title} {Spiral
  superconducting nanowire single-photon detector with efficiency over 50\% at
  1550 nm wavelength},\ }\href@noop {} {\bibfield  {journal} {\bibinfo
  {journal} {Superconductor Science and Technology}\ }\textbf {\bibinfo
  {volume} {30}},\ \bibinfo {pages} {074004} (\bibinfo {year}
  {2017})}\BibitemShut {NoStop}%
\bibitem [{\citenamefont {Verma}\ \emph {et~al.}(2012)\citenamefont {Verma},
  \citenamefont {Marsili}, \citenamefont {Harrington}, \citenamefont {Lita},
  \citenamefont {Mirin},\ and\ \citenamefont
  {Nam}}]{verma_three-dimensional_2012}%
  \BibitemOpen
  \bibfield  {author} {\bibinfo {author} {\bibfnamefont {V.~B.}\ \bibnamefont
  {Verma}}, \bibinfo {author} {\bibfnamefont {F.}~\bibnamefont {Marsili}},
  \bibinfo {author} {\bibfnamefont {S.}~\bibnamefont {Harrington}}, \bibinfo
  {author} {\bibfnamefont {A.~E.}\ \bibnamefont {Lita}}, \bibinfo {author}
  {\bibfnamefont {R.~P.}\ \bibnamefont {Mirin}},\ and\ \bibinfo {author}
  {\bibfnamefont {S.~W.}\ \bibnamefont {Nam}},\ }\bibfield  {title} {\bibinfo
  {title} {A three-dimensional, polarization-insensitive superconducting
  nanowire avalanche photodetector},\ }\href@noop {} {\bibfield  {journal}
  {\bibinfo  {journal} {Appl. Phys. Lett.}\ }\textbf {\bibinfo {volume}
  {101}},\ \bibinfo {pages} {251114} (\bibinfo {year} {2012})}\BibitemShut
  {NoStop}%
\bibitem [{\citenamefont {Xu}\ \emph {et~al.}(2017)\citenamefont {Xu},
  \citenamefont {Zheng}, \citenamefont {Qin}, \citenamefont {Yan},
  \citenamefont {Zhu}, \citenamefont {Kang}, \citenamefont {Zhang},
  \citenamefont {Jia}, \citenamefont {Tu}, \citenamefont {Jin}, \citenamefont
  {Xu}, \citenamefont {Chen},\ and\ \citenamefont
  {Wu}}]{xu_demonstration_2017}%
  \BibitemOpen
  \bibfield  {author} {\bibinfo {author} {\bibfnamefont {R.}~\bibnamefont
  {Xu}}, \bibinfo {author} {\bibfnamefont {F.}~\bibnamefont {Zheng}}, \bibinfo
  {author} {\bibfnamefont {D.}~\bibnamefont {Qin}}, \bibinfo {author}
  {\bibfnamefont {X.}~\bibnamefont {Yan}}, \bibinfo {author} {\bibfnamefont
  {G.}~\bibnamefont {Zhu}}, \bibinfo {author} {\bibfnamefont {L.}~\bibnamefont
  {Kang}}, \bibinfo {author} {\bibfnamefont {L.}~\bibnamefont {Zhang}},
  \bibinfo {author} {\bibfnamefont {X.}~\bibnamefont {Jia}}, \bibinfo {author}
  {\bibfnamefont {X.}~\bibnamefont {Tu}}, \bibinfo {author} {\bibfnamefont
  {B.}~\bibnamefont {Jin}}, \bibinfo {author} {\bibfnamefont {W.}~\bibnamefont
  {Xu}}, \bibinfo {author} {\bibfnamefont {J.}~\bibnamefont {Chen}},\ and\
  \bibinfo {author} {\bibfnamefont {P.}~\bibnamefont {Wu}},\ }\bibfield
  {title} {\bibinfo {title} {Demonstration of polarization-insensitive
  superconducting nanowire single-photon detector with si compensation layer},\
  }\href@noop {} {\bibfield  {journal} {\bibinfo  {journal} {J. Light.
  Technol.}\ }\textbf {\bibinfo {volume} {35}},\ \bibinfo {pages} {4707}
  (\bibinfo {year} {2017})}\BibitemShut {NoStop}%
\bibitem [{\citenamefont {Mukhtarova}\ \emph {et~al.}(2018)\citenamefont
  {Mukhtarova}, \citenamefont {Redaelli}, \citenamefont {Hazra}, \citenamefont
  {Machhadani}, \citenamefont {Lequien}, \citenamefont {Hofheinz},
  \citenamefont {Thomassin}, \citenamefont {Gustavo}, \citenamefont {Zichi},
  \citenamefont {Zwiller}, \citenamefont {Monroy},\ and\ \citenamefont
  {Gérard}}]{mukhtarova_polarization-insensitive_2018}%
  \BibitemOpen
  \bibfield  {author} {\bibinfo {author} {\bibfnamefont {A.}~\bibnamefont
  {Mukhtarova}}, \bibinfo {author} {\bibfnamefont {L.}~\bibnamefont
  {Redaelli}}, \bibinfo {author} {\bibfnamefont {D.}~\bibnamefont {Hazra}},
  \bibinfo {author} {\bibfnamefont {H.}~\bibnamefont {Machhadani}}, \bibinfo
  {author} {\bibfnamefont {S.}~\bibnamefont {Lequien}}, \bibinfo {author}
  {\bibfnamefont {M.}~\bibnamefont {Hofheinz}}, \bibinfo {author}
  {\bibfnamefont {J.}~\bibnamefont {Thomassin}}, \bibinfo {author}
  {\bibfnamefont {F.}~\bibnamefont {Gustavo}}, \bibinfo {author} {\bibfnamefont
  {J.}~\bibnamefont {Zichi}}, \bibinfo {author} {\bibfnamefont
  {V.}~\bibnamefont {Zwiller}}, \bibinfo {author} {\bibfnamefont
  {E.}~\bibnamefont {Monroy}},\ and\ \bibinfo {author} {\bibfnamefont {J.-M.}\
  \bibnamefont {Gérard}},\ }\bibfield  {title} {\bibinfo {title}
  {Polarization-insensitive fiber-coupled superconducting-nanowire single
  photon detector using a high-index dielectric capping layer},\ }\href@noop {}
  {\bibfield  {journal} {\bibinfo  {journal} {Opt. Express}\ }\textbf {\bibinfo
  {volume} {26}},\ \bibinfo {pages} {17697} (\bibinfo {year}
  {2018})}\BibitemShut {NoStop}%
\bibitem [{\citenamefont {{Verma}}\ \emph {et~al.}(2015)\citenamefont
  {{Verma}}, \citenamefont {{Korzh}}, \citenamefont {{Bussieres}},
  \citenamefont {{Horansky}}, \citenamefont {{Dyer}}, \citenamefont {{Lita}},
  \citenamefont {{Vayshenker}}, \citenamefont {{Marsili}}, \citenamefont
  {{Shaw}}, \citenamefont {{Zbinden}}, \citenamefont {{Mirin}},\ and\
  \citenamefont {{Nam}}}]{verma_high-efficiency_2015}%
  \BibitemOpen
  \bibfield  {author} {\bibinfo {author} {\bibfnamefont {V.~B.}\ \bibnamefont
  {{Verma}}}, \bibinfo {author} {\bibfnamefont {B.}~\bibnamefont {{Korzh}}},
  \bibinfo {author} {\bibfnamefont {F.}~\bibnamefont {{Bussieres}}}, \bibinfo
  {author} {\bibfnamefont {R.~D.}\ \bibnamefont {{Horansky}}}, \bibinfo
  {author} {\bibfnamefont {S.~D.}\ \bibnamefont {{Dyer}}}, \bibinfo {author}
  {\bibfnamefont {A.~E.}\ \bibnamefont {{Lita}}}, \bibinfo {author}
  {\bibfnamefont {I.}~\bibnamefont {{Vayshenker}}}, \bibinfo {author}
  {\bibfnamefont {F.}~\bibnamefont {{Marsili}}}, \bibinfo {author}
  {\bibfnamefont {M.~D.}\ \bibnamefont {{Shaw}}}, \bibinfo {author}
  {\bibfnamefont {H.}~\bibnamefont {{Zbinden}}}, \bibinfo {author}
  {\bibfnamefont {R.~P.}\ \bibnamefont {{Mirin}}},\ and\ \bibinfo {author}
  {\bibfnamefont {S.~W.}\ \bibnamefont {{Nam}}},\ }\bibfield  {title} {\bibinfo
  {title} {High-efficiency superconducting nanowire single-photon detectors
  fabricated from mosi thin-films},\ }\href@noop {} {\bibfield  {journal}
  {\bibinfo  {journal} {Opt. Express}\ }\textbf {\bibinfo {volume} {23}},\
  \bibinfo {pages} {33792} (\bibinfo {year} {2015})}\BibitemShut {NoStop}%
\bibitem [{\citenamefont {Gu}\ \emph {et~al.}()\citenamefont {Gu},
  \citenamefont {Cheng}, \citenamefont {Zhu},\ and\ \citenamefont
  {Hu}}]{Gu_fractal_2015}%
  \BibitemOpen
  \bibfield  {author} {\bibinfo {author} {\bibfnamefont {C.}~\bibnamefont
  {Gu}}, \bibinfo {author} {\bibfnamefont {Y.}~\bibnamefont {Cheng}}, \bibinfo
  {author} {\bibfnamefont {X.}~\bibnamefont {Zhu}},\ and\ \bibinfo {author}
  {\bibfnamefont {X.}~\bibnamefont {Hu}},\ }\bibfield  {title} {\bibinfo
  {title} {Fractal-inspired, polarization-insensitive superconducting nanowire
  single-photon detectors,},\ }\href@noop {} {\bibfield  {journal} {\bibinfo
  {journal} {\textit{Novel Optical Materials and Applications} (Optical Society
  of America, 2015)}\ ,\ \bibinfo {pages} {paper JM3A.10}}}\bibinfo {note}
  {DOI: 10.1364/IPRSN.2015.JM3A.10}\BibitemShut {NoStop}%
\bibitem [{\citenamefont {Chi}\ \emph {et~al.}(2018)\citenamefont {Chi},
  \citenamefont {Zou}, \citenamefont {Gu}, \citenamefont {Zichi}, \citenamefont
  {Cheng}, \citenamefont {Hu}, \citenamefont {Lan}, \citenamefont {Chen},
  \citenamefont {Lin}, \citenamefont {Zwiller},\ and\ \citenamefont
  {Hu}}]{chi_fractal_2018}%
  \BibitemOpen
  \bibfield  {author} {\bibinfo {author} {\bibfnamefont {X.}~\bibnamefont
  {Chi}}, \bibinfo {author} {\bibfnamefont {K.}~\bibnamefont {Zou}}, \bibinfo
  {author} {\bibfnamefont {C.}~\bibnamefont {Gu}}, \bibinfo {author}
  {\bibfnamefont {J.}~\bibnamefont {Zichi}}, \bibinfo {author} {\bibfnamefont
  {Y.}~\bibnamefont {Cheng}}, \bibinfo {author} {\bibfnamefont
  {N.}~\bibnamefont {Hu}}, \bibinfo {author} {\bibfnamefont {X.}~\bibnamefont
  {Lan}}, \bibinfo {author} {\bibfnamefont {S.}~\bibnamefont {Chen}}, \bibinfo
  {author} {\bibfnamefont {Z.}~\bibnamefont {Lin}}, \bibinfo {author}
  {\bibfnamefont {V.}~\bibnamefont {Zwiller}},\ and\ \bibinfo {author}
  {\bibfnamefont {X.}~\bibnamefont {Hu}},\ }\bibfield  {title} {\bibinfo
  {title} {Fractal superconducting nanowire single-photon detectors with
  reduced polarization sensitivity},\ }\href@noop {} {\bibfield  {journal}
  {\bibinfo  {journal} {Opt. Lett.}\ }\textbf {\bibinfo {volume} {43}},\
  \bibinfo {pages} {5017} (\bibinfo {year} {2018})}\BibitemShut {NoStop}%
\bibitem [{PSn()}]{PSnote}%
  \BibitemOpen
  \href@noop {} {}\bibinfo {note} {We note that some researchers define PS as
  $(\rm{SDE}_{\rm{max}}-\rm{SDE}_{\rm{min}})/(\rm{SDE}_{\rm{max}}+\rm{SDE}_{\rm{min}})$,
  for example, in Refs. 30 and 34, which can be equivalently used for
  characterizing the polarization dependence of the detection
  efficiency}\BibitemShut {NoStop}%
\bibitem [{\citenamefont {Clem}\ and\ \citenamefont
  {Berggren}(2011)}]{clem_geometry-dependent_2011}%
  \BibitemOpen
  \bibfield  {author} {\bibinfo {author} {\bibfnamefont {J.~R.}\ \bibnamefont
  {Clem}}\ and\ \bibinfo {author} {\bibfnamefont {K.~K.}\ \bibnamefont
  {Berggren}},\ }\bibfield  {title} {\bibinfo {title} {Geometry-dependent
  critical currents in superconducting nanocircuits},\ }\href@noop {}
  {\bibfield  {journal} {\bibinfo  {journal} {Phys. Rev. B}\ }\textbf {\bibinfo
  {volume} {84}},\ \bibinfo {pages} {174510} (\bibinfo {year}
  {2011})}\BibitemShut {NoStop}%
\bibitem [{\citenamefont {Su}\ \emph {et~al.}(2008)\citenamefont {Su},
  \citenamefont {Wang}, \citenamefont {Huang},\ and\ \citenamefont
  {Lancaster}}]{su_superconducting_2008}%
  \BibitemOpen
  \bibfield  {author} {\bibinfo {author} {\bibfnamefont {H.~T.}\ \bibnamefont
  {Su}}, \bibinfo {author} {\bibfnamefont {Y.}~\bibnamefont {Wang}}, \bibinfo
  {author} {\bibfnamefont {F.}~\bibnamefont {Huang}},\ and\ \bibinfo {author}
  {\bibfnamefont {M.~J.}\ \bibnamefont {Lancaster}},\ }\bibfield  {title}
  {\bibinfo {title} {Superconducting delay lines},\ }\href@noop {} {\bibfield
  {journal} {\bibinfo  {journal} {J. Supercond. Novel Magn.}\ }\textbf
  {\bibinfo {volume} {21}},\ \bibinfo {pages} {7} (\bibinfo {year}
  {2008})}\BibitemShut {NoStop}%
\bibitem [{\citenamefont {Zichi}\ \emph {et~al.}(2019)\citenamefont {Zichi},
  \citenamefont {Chang}, \citenamefont {Steinhauer}, \citenamefont {von
  Fieandt}, \citenamefont {Los}, \citenamefont {Visser}, \citenamefont
  {Kalhor}, \citenamefont {Lettner}, \citenamefont {Elshaari}, \citenamefont
  {Esmaeil~Zadeh},\ and\ \citenamefont {Zwiller}}]{zichi_optimizing_2019}%
  \BibitemOpen
  \bibfield  {author} {\bibinfo {author} {\bibfnamefont {J.}~\bibnamefont
  {Zichi}}, \bibinfo {author} {\bibfnamefont {J.}~\bibnamefont {Chang}},
  \bibinfo {author} {\bibfnamefont {S.}~\bibnamefont {Steinhauer}}, \bibinfo
  {author} {\bibfnamefont {K.}~\bibnamefont {von Fieandt}}, \bibinfo {author}
  {\bibfnamefont {J.~W.~N.}\ \bibnamefont {Los}}, \bibinfo {author}
  {\bibfnamefont {G.}~\bibnamefont {Visser}}, \bibinfo {author} {\bibfnamefont
  {N.}~\bibnamefont {Kalhor}}, \bibinfo {author} {\bibfnamefont
  {T.}~\bibnamefont {Lettner}}, \bibinfo {author} {\bibfnamefont {A.~W.}\
  \bibnamefont {Elshaari}}, \bibinfo {author} {\bibfnamefont {I.}~\bibnamefont
  {Esmaeil~Zadeh}},\ and\ \bibinfo {author} {\bibfnamefont {V.}~\bibnamefont
  {Zwiller}},\ }\bibfield  {title} {\bibinfo {title} {Optimizing the
  stoichiometry of ultrathin {NbTiN} films for high-performance superconducting
  nanowire single-photon detectors},\ }\href@noop {} {\bibfield  {journal}
  {\bibinfo  {journal} {Opt. Express}\ }\textbf {\bibinfo {volume} {27}},\
  \bibinfo {pages} {26579} (\bibinfo {year} {2019})}\BibitemShut {NoStop}%
\bibitem [{\citenamefont {Miller}\ \emph {et~al.}(2011)\citenamefont {Miller},
  \citenamefont {Lita}, \citenamefont {Calkins}, \citenamefont {Vayshenker},
  \citenamefont {Gruber},\ and\ \citenamefont {Nam}}]{miller_compact_2011}%
  \BibitemOpen
  \bibfield  {author} {\bibinfo {author} {\bibfnamefont {A.~J.}\ \bibnamefont
  {Miller}}, \bibinfo {author} {\bibfnamefont {A.~E.}\ \bibnamefont {Lita}},
  \bibinfo {author} {\bibfnamefont {B.}~\bibnamefont {Calkins}}, \bibinfo
  {author} {\bibfnamefont {I.}~\bibnamefont {Vayshenker}}, \bibinfo {author}
  {\bibfnamefont {S.~M.}\ \bibnamefont {Gruber}},\ and\ \bibinfo {author}
  {\bibfnamefont {S.~W.}\ \bibnamefont {Nam}},\ }\bibfield  {title} {\bibinfo
  {title} {Compact cryogenic self-aligning fiber-to-detector coupling with
  losses below one percent},\ }\href@noop {} {\bibfield  {journal} {\bibinfo
  {journal} {Opt. Express}\ }\textbf {\bibinfo {volume} {19}},\ \bibinfo
  {pages} {9102} (\bibinfo {year} {2011})}\BibitemShut {NoStop}%
\bibitem [{\citenamefont {{Miki}}\ \emph {et~al.}(2017)\citenamefont {{Miki}},
  \citenamefont {{Yabuno}}, \citenamefont {{Yamashita}},\ and\ \citenamefont
  {{Terai}}}]{miki2017stable}%
  \BibitemOpen
  \bibfield  {author} {\bibinfo {author} {\bibfnamefont {S.}~\bibnamefont
  {{Miki}}}, \bibinfo {author} {\bibfnamefont {M.}~\bibnamefont {{Yabuno}}},
  \bibinfo {author} {\bibfnamefont {T.}~\bibnamefont {{Yamashita}}},\ and\
  \bibinfo {author} {\bibfnamefont {H.}~\bibnamefont {{Terai}}},\ }\bibfield
  {title} {\bibinfo {title} {Stable, high-performance operation of a
  fiber-coupled superconducting nanowire avalanche photon detector.},\
  }\href@noop {} {\bibfield  {journal} {\bibinfo  {journal} {Opt. Express}\
  }\textbf {\bibinfo {volume} {25}},\ \bibinfo {pages} {6796} (\bibinfo {year}
  {2017})}\BibitemShut {NoStop}%
\bibitem [{\citenamefont {{Chen}}\ \emph {et~al.}(2015)\citenamefont {{Chen}},
  \citenamefont {{You}}, \citenamefont {{Zhang}}, \citenamefont {{Yang}},
  \citenamefont {{Li}}, \citenamefont {{Zhang}}, \citenamefont {{Wang}},\ and\
  \citenamefont {{Xie}}}]{chen2015dark}%
  \BibitemOpen
  \bibfield  {author} {\bibinfo {author} {\bibfnamefont {S.}~\bibnamefont
  {{Chen}}}, \bibinfo {author} {\bibfnamefont {L.}~\bibnamefont {{You}}},
  \bibinfo {author} {\bibfnamefont {W.}~\bibnamefont {{Zhang}}}, \bibinfo
  {author} {\bibfnamefont {X.}~\bibnamefont {{Yang}}}, \bibinfo {author}
  {\bibfnamefont {H.}~\bibnamefont {{Li}}}, \bibinfo {author} {\bibfnamefont
  {L.}~\bibnamefont {{Zhang}}}, \bibinfo {author} {\bibfnamefont
  {Z.}~\bibnamefont {{Wang}}},\ and\ \bibinfo {author} {\bibfnamefont
  {X.}~\bibnamefont {{Xie}}},\ }\bibfield  {title} {\bibinfo {title} {Dark
  counts of superconducting nanowire single-photon detector under
  illumination.},\ }\href@noop {} {\bibfield  {journal} {\bibinfo  {journal}
  {Opt. Express}\ }\textbf {\bibinfo {volume} {23}},\ \bibinfo {pages} {10786}
  (\bibinfo {year} {2015})}\BibitemShut {NoStop}%
\bibitem [{\citenamefont {Gerrits}\ \emph {et~al.}(2019)\citenamefont
  {Gerrits}, \citenamefont {Migdall}, \citenamefont {Bienfang}, \citenamefont
  {Lehman}, \citenamefont {Nam}, \citenamefont {Splett}, \citenamefont
  {Vayshenker},\ and\ \citenamefont {Wang}}]{gerrits2019calibration}%
  \BibitemOpen
  \bibfield  {author} {\bibinfo {author} {\bibfnamefont {T.}~\bibnamefont
  {Gerrits}}, \bibinfo {author} {\bibfnamefont {A.}~\bibnamefont {Migdall}},
  \bibinfo {author} {\bibfnamefont {J.~C.}\ \bibnamefont {Bienfang}}, \bibinfo
  {author} {\bibfnamefont {J.}~\bibnamefont {Lehman}}, \bibinfo {author}
  {\bibfnamefont {S.~W.}\ \bibnamefont {Nam}}, \bibinfo {author} {\bibfnamefont
  {J.}~\bibnamefont {Splett}}, \bibinfo {author} {\bibfnamefont
  {I.}~\bibnamefont {Vayshenker}},\ and\ \bibinfo {author} {\bibfnamefont
  {J.}~\bibnamefont {Wang}},\ }\bibfield  {title} {\bibinfo {title}
  {Calibration of free-space and fiber-coupled single-photon detectors},\
  }\href@noop {} {\bibfield  {journal} {\bibinfo  {journal} {Metrologia}\
  }\textbf {\bibinfo {volume} {57}},\ \bibinfo {pages} {015002} (\bibinfo
  {year} {2019})}\BibitemShut {NoStop}%
\bibitem [{\citenamefont {{Marsili}}\ \emph {et~al.}(2012)\citenamefont
  {{Marsili}}, \citenamefont {{Najafi}}, \citenamefont {{Dauler}},
  \citenamefont {{Molnar}},\ and\ \citenamefont
  {{Berggren}}}]{marsili2012afterpulsing}%
  \BibitemOpen
  \bibfield  {author} {\bibinfo {author} {\bibfnamefont {F.}~\bibnamefont
  {{Marsili}}}, \bibinfo {author} {\bibfnamefont {F.}~\bibnamefont {{Najafi}}},
  \bibinfo {author} {\bibfnamefont {E.}~\bibnamefont {{Dauler}}}, \bibinfo
  {author} {\bibfnamefont {R.~J.}\ \bibnamefont {{Molnar}}},\ and\ \bibinfo
  {author} {\bibfnamefont {K.~K.}\ \bibnamefont {{Berggren}}},\ }\bibfield
  {title} {\bibinfo {title} {Afterpulsing and instability in superconducting
  nanowire avalanche photodetectors},\ }\href@noop {} {\bibfield  {journal}
  {\bibinfo  {journal} {Appl. Phys. Lett.}\ }\textbf {\bibinfo {volume}
  {100}},\ \bibinfo {pages} {112601} (\bibinfo {year} {2012})}\BibitemShut
  {NoStop}%
\bibitem [{\citenamefont {Sidorova}\ \emph {et~al.}(2017)\citenamefont
  {Sidorova}, \citenamefont {Semenov}, \citenamefont {H{\"u}bers},
  \citenamefont {Charaev}, \citenamefont {Kuzmin}, \citenamefont {Doerner},\
  and\ \citenamefont {Siegel}}]{sidorova2017physical}%
  \BibitemOpen
  \bibfield  {author} {\bibinfo {author} {\bibfnamefont {M.}~\bibnamefont
  {Sidorova}}, \bibinfo {author} {\bibfnamefont {A.}~\bibnamefont {Semenov}},
  \bibinfo {author} {\bibfnamefont {H.-W.}\ \bibnamefont {H{\"u}bers}},
  \bibinfo {author} {\bibfnamefont {I.}~\bibnamefont {Charaev}}, \bibinfo
  {author} {\bibfnamefont {A.}~\bibnamefont {Kuzmin}}, \bibinfo {author}
  {\bibfnamefont {S.}~\bibnamefont {Doerner}},\ and\ \bibinfo {author}
  {\bibfnamefont {M.}~\bibnamefont {Siegel}},\ }\bibfield  {title} {\bibinfo
  {title} {Physical mechanisms of timing jitter in photon detection by
  current-carrying superconducting nanowires},\ }\href@noop {} {\bibfield
  {journal} {\bibinfo  {journal} {Phys. Rev. B}\ }\textbf {\bibinfo {volume}
  {96}},\ \bibinfo {pages} {184504} (\bibinfo {year} {2017})}\BibitemShut
  {NoStop}%
\bibitem [{\citenamefont {Palik}(1985)}]{palik1998handbook}%
  \BibitemOpen
  \bibfield  {author} {\bibinfo {author} {\bibfnamefont {E.~D.}\ \bibnamefont
  {Palik}},\ }\href@noop {} {\emph {\bibinfo {title} {Handbook of optical
  constants of solids}}}\ (\bibinfo  {publisher} {Academic Press},\ \bibinfo
  {address} {Boston},\ \bibinfo {year} {1985})\BibitemShut {NoStop}%
\bibitem [{hu2()}]{hu2011efficient}%
  \BibitemOpen
  \href@noop {} {}\bibinfo {note} {X. Hu, Efficient superconducting-nanowire
  single-photondetectors and their applications in quantum optics. Ph.D.
  thesis, Massachusetts Institute of Technology (2011).}\BibitemShut {Stop}%
\bibitem [{\citenamefont {Bright}\ \emph {et~al.}(2013)\citenamefont {Bright},
  \citenamefont {Watjen}, \citenamefont {Zhang}, \citenamefont {Muratore},
  \citenamefont {Voevodin}, \citenamefont {Koukis}, \citenamefont {Tanner},\
  and\ \citenamefont {Arenas}}]{bright2013infrared}%
  \BibitemOpen
  \bibfield  {author} {\bibinfo {author} {\bibfnamefont {T.~J.}\ \bibnamefont
  {Bright}}, \bibinfo {author} {\bibfnamefont {J.}~\bibnamefont {Watjen}},
  \bibinfo {author} {\bibfnamefont {Z.}~\bibnamefont {Zhang}}, \bibinfo
  {author} {\bibfnamefont {C.}~\bibnamefont {Muratore}}, \bibinfo {author}
  {\bibfnamefont {A.~A.}\ \bibnamefont {Voevodin}}, \bibinfo {author}
  {\bibfnamefont {D.}~\bibnamefont {Koukis}}, \bibinfo {author} {\bibfnamefont
  {D.~B.}\ \bibnamefont {Tanner}},\ and\ \bibinfo {author} {\bibfnamefont
  {D.~J.}\ \bibnamefont {Arenas}},\ }\bibfield  {title} {\bibinfo {title}
  {Infrared optical properties of amorphous and nanocrystalline ta$_2$o$_5$
  thin films},\ }\href@noop {} {\bibfield  {journal} {\bibinfo  {journal}
  {Journal of Applied Physics}\ }\textbf {\bibinfo {volume} {114}},\ \bibinfo
  {pages} {083515} (\bibinfo {year} {2013})}\BibitemShut {NoStop}%
\end{thebibliography}%

%%%%%%%%%%%%%%%%%%%%%%%%%%%%%%%%%%%%%%%%%%%%%%%%%%%%%%%%%%%%%%%%%%%%%%%%   end

% The \nocite command causes all entries in a bibliography to be printed out
% whether or not they are actually referenced in the text. This is appropriate
% for the sample file to show the different styles of references, but authors
% most likely will not want to use it.
%\nocite{*}
% Produces the bibliography via BibTeX.
%\bibliographyfullrefs{Exported Items_t1}
\end{document}